\documentclass[aps,10pt,twocolumn,nofootinbib]{revtex4-1}

\usepackage{amsmath}
\usepackage{mathrsfs}
\usepackage{hyperref}
\usepackage{tikz}
\usepackage{orcidlink}
\usepackage{txfonts}

\newcommand{\ud}{\mathrm{d}}
\newcommand{\RR}{\mathbb{R}}
\newcommand{\scE}{\mathscr{E}}
\newcommand{\scL}{\mathscr{L}}
\newcommand{\mcM}{\mathcal{M}}
\newcommand{\mcN}{\mathcal{N}}
\newcommand{\mcT}{\mathcal{T}}
\newcommand{\mcC}{\mathcal{C}}
\newcommand{\mcP}{\mathcal{P}}
\newcommand{\fK}{\mathfrak{K}}
\newcommand{\fQ}{\mathfrak{Q}}

\newcommand{\fS}{\mathfrak{S}}

\begin{document}

\title{On the integrability of extended test body dynamics around black holes}

\author{Paul Ramond,\,
\orcidlink{0000-0001-7123-0039}}
\email{paul.ramond@obspm.fr}
\affiliation{IMCCE, Observatoire de Paris, Université PSL, 77 Avenue Denfert-Rochereau, FR-75014, Paris}

\begin{abstract}
In general relativity, the motion of an extended test body is influenced by its proper rotation, or \textit{spin}. We present a covariant and physically self-consistent Hamiltonian framework to study this motion, up to quadratic order in the body's spin, including a spin-induced quadrupole, and in an arbitrary background spacetime. The choice of spin supplementary condition and degeneracies associated with local Lorentz invariance are treated rigorously with adapted tools from Hamiltonian mechanics. Applying the formalism to a background space-time described by the Kerr metric, we prove that the motion of \textit{any} test compact object around a rotating black hole defines an \textit{integrable} Hamiltonian system to \textit{linear} order in the body's spin. Moreover, this integrability still holds at \textit{quadratic} order in spin when the compact object has the deformability expected for an isolated \textit{black hole}. By exploiting the unique symmetries at play in black hole binaries, our analytical results clarify longstanding numerical conjectures regarding spin-induced chaos in the motion of asymmetric compact binaries, and may provide a powerful framework to improve current gravitational waveform modelling.

\end{abstract}

\maketitle

\textit{Introduction.---} 
The motion of a celestial body in a given gravitational field is, arguably, the oldest and most fundamental problem in theoretical astrophysics. Although it generally resists simple, analytical solutions – mostly due to the complex internal composition of celestial bodies – the bulk motion of such a body adheres to conservation laws for linear and angular momentum. The microscopic details enter the dynamics at finer levels of description. This remarkable feature is consequential to the \textit{universality} of gravitation, the core of Einstein’s theory of general relativity. The natural setup to understand these properties is through general relativistic multipolar expansions of a body \cite{Ma.37,Di.64,Ha.15,MaPoWa.22}. These schemes assert that, aside from dissipation, the motion of an extended body in a given spacetime mirrors the worldline of a point object endowed with multipoles. At \textit{monopolar} order, the body is replaced by a point mass whose worldline is a geodesic of the background spacetime: A mere rephrasing of Einstein's geodesic principle \cite{Einstein.14}. At \textit{dipolar} order, the body’s proper rotation (hereafter \textit{spin}) couples to spacetime’s curvature, thus deviating its worldline from a geodesic. Remarkably, this dipolar, non-geodesic motion is still universal: The body’s momenta obey purely \textit{kinematical} evolution equations along a worldline that is completely determined by the background geometry \cite{Ha.15}. Universality disappears at \textit{quadrupolar} order, with body-specific multipoles entering the description and \textit{dynamically} driving the evolution equations via self-forces and torques \cite{HaDw.23}. 

\vspace{1mm}The framework of Hamiltonian mechanics, originally designed specifically for conservative systems, naturally emerges as a fundamental tool to study these multipolar evolution equations. Hamiltonian treatments have been utilized in nearly all analytical approximation schemes developed to address the general-relativistic two-body problem \cite{DaNa,Fu.al.17,Schafer.al.18}. In the case of simple or highly symmetric systems, Hamilton's equations can be solved by \textit{quadrature}, i.e., involving straightforward algebraic manipulations and one-dimensional integrals. Such systems, referred to as \textit{integrable}, are invaluable to study the dynamics with powerful and unique mathematical tools. Generic Hamiltonian systems have no inherent reason to be integrable and will most often exhibit chaotic dynamics \cite{Poin}. Nevertheless, in 1968, B.~Carter was able to prove that the relativistic motion of a monopolar body around a rotating black hole, as described by geodesics of the Kerr metric, was integrable \cite{Carter.68}. 
The new integral of motion found by Carter was then shown to be connected to a very important geometric object, the Killing-Yano tensor of the Kerr metric \cite{WaPe.70,Floyd.73}, which, alone, generates all the Kerr symmetries that lead to geodesic integrability \cite{HuSo.73}. 

\vspace{1mm}Naturally, the integrability of Kerr geodesics has prompted many to question the integrability around black holes at higher multipolar order, where, in particular, the body's spin breaks geodesic motion and Carter's constant is not conserved anymore. To speak of integrability, one requires a Hamiltonian formulation of the dynamics beyond geodesics. Several have been proposed %\footnote{Ref \cite{ViKuStHi.16} contains, to our knowledge, the only consistent quadratic-in-spin Hamiltonian framework for test bodies in a background spacetime. At this order, different spin supplementary condition may describe different physical systems (cf. Sec.~\cite{ViKuStHi.16}). The equivalence of their framework and ours is, therefore, far from evident and worst the investigation.}
\cite{Ba.al.09,Stei.11,dAKuvHo.15,ViKuStHi.16,WiStLu.19} for other objectives than showing integrability. However, most of these frameworks are limited to linear-in-spin dynamics, and either lack general covariance, work with degenerate phase spaces with too many dimensions and/or implement the (crucial) spin supplementary condition heuristically. While not intrinsically problematic for specific goals, these aforementioned features limit one's ability to apply standard tools of classical Hamiltonian theory, thereby resulting in debatable conclusions about integrability (or lack thereof) \cite{KuLeLuSe.16,WitzHJ.19,ComDru.22,GrMo.23}. Supported by numerical studies of the revealing signatures of chaotic motion \cite{SuMa.97,SuMa.99,Hartl.03,Hartl2.03,KaoCho.05,Han.08,KuLeLuSe.16,Lukes.18,WiStLu.19}, this has led to the current conjecture that a body's spin should generally break integrability around rotating black holes, even at linear order in spin. 

In this work, we propose to go beyond the pre-existing numerical and/or conjectural results, and answer analytically whether the body's spin breaks Carter's geodesic integrability in a Kerr background. 
Reaching this goal up to quadratic order in the body's spin, we found, required to step \textit{aside} from common practices, such as the use of (i) Marck tetrads \cite{Marck.83,VdM.20} (limited to the linear-in-spin dynamics and requiring the introduction of extra fiducial parameters in the system \cite{WitzHJ.19}); (ii) 3+1 spacetime decomposition and the Newton-Wigner supplementary condition \cite{KuLeLuSe.16,ViKuStHi.16} (which break general covariance and lead to a pure-constraint Hamiltonian via singular Legendre transform); and (iii) the use of too large a phase space and/or with degenerate Poisson structure on it \cite{ComDru.22} (which leads to a non-integrable, but also non-physical system, at least in Kerr). All these aforementioned methods, and their (sometimes) unpractical consequences, are absent from our framework. On the contrary, everything that follows relies solely on covariant calculations and tools from Poisson geometry \cite{Vaisman.12}.

Our results can be summarized as the outcomes of a three-steps process. First, we describe the quadratic-in-spin dynamics of an extended test body in \textit{any} spacetime as a Hamiltonian system that is at once covariant, non-degenerate, on a 10-dimensional phase space and does not require any heuristic or external ad-hoc parameter-fixing. Second, when the framework is applied to a background describing a Kerr black hole, the resulting system is proven to be \textit{integrable} at \textit{linear} order in spin, owing to the existence of enough independent integrals of motion for \textit{any} test body. Third, still in a Kerr background, we show that quadratic-in-spin terms (from both dipole effects and a spin-induced quadrupole) \textit{do not} break integrability when the test object has the quadrupole structure expected for an isolated black hole. Possible applications of the results are then discussed.

Geometrical units $G=c=1$ are used throughout, and our conventions for Lorentzian and Poisson geometry follow \cite{Wald} and \cite{Arn}, respectively. %Extended discussions and detailed calculations are presented in \cite{Ra.PapI.24} and \cite{Ra.al.quad.24} for linear and quadratic order in spin, respectively. 

\vspace{1mm}{\it Quadrupolar dynamics.---} We consider a fixed background spacetime $(\scE,g_{ab})$ where $\scE$ is a 4-dimensional (4D) manifold covered with coordinates $x^\alpha$, and $g_{ab}$ is a metric tensor on $\scE$. Following classical multipolar expansion schemes \cite{Di.74,Ha.15}, the motion of an extended body in $\scE$ can be effectively described by the trajectory of a representative point particle endowed with a number of multipoles. Let $\scL\subset\scE$ be that particle's worldline, $\tau$ its proper time and $u^a$ its four-velocity, normalized as $g_{ab} u^a u^b=-1$. The multipolar expansion accounts for both translational and rotational degrees of freedom of the body, encoded into a momentum 1-form $p_a$ and an antisymmetric spin tensor $S^{ab}$. The evolution equations for $(p_a,S^{ab})$ along $\scL$ are known as the
Mathisson-Papapetrou-Tulczyjew-Dixon (MPTD) equations 
\cite{Ma.37,Pa.51,Tu.59,Di.74,Ha.12} 
\begin{equation} \label{EE}
\nabla_u p_a = R_{abcd} S^{bc} u^d +F_a \,, \quad \nabla_u S^{ab} = 2 p^{[a} u^{b]} +N^{ab} \,,
\end{equation}
where $p^a:=g^{ab}p_b$, $\nabla$ is the $g_{ab}$-compatible connection, $\nabla_u:=u^a\nabla_a$ is the covariant derivative along $\scL$ and $R_{abcd}$ is the Riemann curvature tensor of $(\scE,g_{ab})$. 
In this work, we truncate the multipolar expansion to quadrupolar order. The force and torque in \eqref{EE} are then given by $F_a=-\tfrac{1}{6}J^{bcde}\nabla_a R_{bcde}$ and $N^{ab}=-\tfrac{4}{3} J^{[a}_{\phantom{a}cde}R^{b]cde}$ \cite{Di.74,Ha.20}, where $J^{abcd}$ is the body's \textit{quadrupole} tensor, whose algebraic symmetries match those of the Riemann tensor. From $(p_a,S^{ab})$, we define the body's (dynamical) mass $\mu^2:=-p_a p^a$ and two spin norms $S_\circ^2:=\tfrac{1}{2}S_{ab}S^{ab}$ and $S_\star^2:=\tfrac{1}{8}\varepsilon_{abcd}S^{ab}S^{cd}$, where $\varepsilon_{abcd}$ is the Levi-Civita tensor associated with $g_{ab}$. In this work, we will consider a spin-induced quadrupole \cite{StPu.12,Ma.15,ViKuStHi.16} 
\begin{equation} \label{Jabcd}
    J^{abcd}:=\frac{3\kappa}{\mu} \bar{p}^{[a}\Theta^{b][c}\bar{p}^{d]} \,, \,\, \text{with} \,\, \begin{cases}
		\, \Theta^{ab}:=g_{cd}S^{ac}S^{db} \, , \\
		\,\,\bar{p}^a := p^a/\mu  \,,
	\end{cases}
\end{equation}
and where $\kappa$ is a dimensionless coupling parameter that encodes the deformability of the extended object. Its value is expected to be $1$ for an isolated Kerr black hole \cite{BiPo.09}, and $\kappa>1$ for other compact object, larger values of $\kappa$ indicating ``stiffer'' equations of state \cite{LaPo.99,Bosh.al.12,Uchi.al.16}. The system \eqref{EE}--\eqref{Jabcd} suffers from two drawbacks. First, it is physically inconsistent: Cubic-in-$S^{ab}$ terms are known to arise at the octupolar order \cite{Ma.15} and contribute to the right-hand sides of Eqs.~\eqref{EE}. Therefore, any spin-induced quadrupolar model is only self-consistent at quadratic order in $S^{ab}$. Second, the system is mathematically ill-posed: For a given spacetime geometry, it is equivalent to 10 ordinary differential equations for 13 unknowns in $(u^a,p_a,S^{ab})$. These issues are solved in two steps, as follows. First, as is customary, we start by closing the differential system \eqref{EE} with four algebraic equations known as a \textit{spin supplementary condition} (SSC) \cite{CoNa.15}. For well-motivated reasons exposed in \cite{Ra.PapI.24}, we choose the \textit{Tulczyjew} SSC:
\begin{equation}\label{TDSSC}
C^b := p_aS^{ab}=0 \,.
\end{equation}
Physically, Eq.~\eqref{TDSSC} is interpreted as the vanishing of the body's mass dipole $\bar{C}^b:=\bar{p}_aS^{ab}$ as measured by an observer of four-velocity $\bar{p}^a$. Second, we combine Eqs.~\eqref{EE} with the covariant derivative of \eqref{TDSSC} to obtain a \textit{momentum-velocity relation} \cite{GraHarWal.10}
\begin{equation} \label{EElin}
    u^a=\bar{p}^a+\frac{1}{2\mu^2} R_{ebcd}S^{eb}\bar{p}^d S^{ac}+\frac{4}{3\mu} J^{[a}_{\phantom{a}cde}R^{b]cde} \bar{p}_b \,,
\end{equation}
where the result has been expanded consistently to quadratic order in $S^{ab}$, for self-consistency of the quadrupolar approximation. This will be always done subsequently. 
Combining Eqs.~\eqref{EE}--\eqref{EElin}, it follows that $S_\star$ vanishes identically \cite{WiStLu.19}, and that both $S_\circ$ and the \textit{effective mass} $\tilde{\mu}:=\mu-\frac{1}{6}R_{abcd}J^{abcd}$ are conserved along $\scL$ at quadratic order in spin \cite{DiI.70,Ma.15,ComDruVin.23}. Projecting Eqs.~\eqref{EE}--\eqref{EElin} onto the coordinate basis $(\partial_\alpha)^a$ and characterizing the worldline in the form $u^\alpha=\ud x^\alpha/\ud \tau$ leads to 14 first order, ordinary differential equations for 14 unknowns $x^\alpha$, $p_\alpha$ and $S^{\alpha\beta}$.

\vspace{1mm}{\it Formulation as a Poisson system.---}
We now wish to turn system \eqref{EE}--\eqref{EElin} into a \textit{Hamiltonian system} in the general sense, i.e., as defined from the three following ingredients: (i) a \textit{phase space $\mcM$}, defined as a $N$-dimensional manifold endowed with $N$ coordinates $y:=(y^1,\ldots,y^N)\in\RR^N$ ; (ii) a \textit{Poisson structure $\Lambda$}, defined as a skew-symmetric $N\times N$ matrix whose entries, denoted $\Lambda^{ij}(y)$, satisfy the Jacobi identity $\Lambda^{\ell(i}\partial_\ell\Lambda^{jk)}=0$, with $\partial_\ell:=\partial/\partial y^\ell$ ; (iii) a \textit{Hamiltonian $H$}, defined as a scalar field $\mcM\rightarrow \RR$.
The triplet $(\mcM,\Lambda,H)$ is called a \textit{Poisson system}. If, for all $y\in\mcM$, the matrix $\Lambda(y)$ has maximal rank, one speaks of a \textit{symplectic} structure. If not, it is said to be \textit{degenerate}. 
Regardless of its degeneracy, the Poisson structure $\Lambda$ defines a \textit{Poisson bracket} via:
\begin{equation} \label{PBnonsymp}
\{F,G\} := \sum_{i,j} \,\Lambda^{ij}(y) \,  \frac{\partial F}{\partial y^i} \, \frac{\partial G}{\partial y^j} \,,
\end{equation}
for any $y$-dependent functions $F,G$. While the geometry of $\mcM$ is fixed by $\Lambda$, the physics rely on a choice of Hamiltonian $H$. The latter defines a preferred set of curves in $\mcM$, solutions to \textit{Hamilton's equation}: $\ud F/\ud \lambda = \{F,H\}$, where $\lambda$ is a parameter uniquely associated with $H$. We now show that system \eqref{EE}--\eqref{EElin} can be written as a Poisson system in $N=14$ dimensions. The phase space $\mcM=\RR^{14}$ is endowed with coordinates $y:=(x^\alpha,p_\alpha,S^{\alpha\beta})\in\RR^{4}\times\RR^{4}\times\RR^{6}$, that coincide physically with their covariant spacetime definitions. The Poisson structure $\Lambda$ is defined uniquely through a given set of Poisson brackets between the coordinates, since Eq.~\eqref{PBnonsymp} implies $\{y^i,y^j\}=\Lambda^{ij}(y)$. Here, the non-vanishing brackets are given by \cite{Souri.70,Kun.72,dAKuvHo.15}
\begin{equation}\label{PBs}
    \begin{split} 
    &\{x^\alpha,p_\beta\} = \delta^\alpha_\beta \,,
    \,\, \{S^{\alpha\beta},S^{\gamma\delta}\} = 2(g^{\alpha[\delta}S^{\gamma]\beta} 
    +g^{\beta[\gamma}S^{\delta]\alpha}) \,,
    \\ 
    &\{p_\alpha,S^{\beta\gamma}\} =  2\Gamma^{[\gamma}_{\delta\alpha} S^{\beta]\delta} \,,
    \,\, \{p_\alpha,p_\beta\} =R_{\alpha\gamma\delta\beta}S^{\gamma\delta}\,,
    \end{split}
\end{equation}
where $\delta^\alpha_\beta$ is the 4D Kronecker symbol. To our knowledge, there does not exist a Hamiltonian in the literature that, when varied with respect to the brackets \eqref{PBs}, generates the covariant, background-independent, quadrupolar MPTD equations \eqref{EE}--\eqref{EElin}. We found such a Hamiltonian by exploiting the conservation of the effective mass $\tilde{\mu}$, which is the sum of the dynamical mass $\mu$ (conserved at linear order in spin, cf.~\cite{Ra.PapI.24}) and a quadrupolar correction. Based on these observations, we construct the natural Hamiltonian $H:=H_g+H_d+H_q$ as that of geodesic motion ($H_g$) with quadratic-in-spin terms due to the dipolar $(H_d)$ and quadrupolar $(H_q)$ sector. Explicitly:
\begin{equation} \label{HTD}
H:=\frac{1}{2}g^{\alpha\beta}p_\alpha p_\beta + R_{\alpha\beta \gamma \delta } \bar{p}^{\alpha} S^{\beta\gamma} \bar{C}^\delta + \frac{\kappa}{2} R_{\alpha \beta \gamma \delta} \bar{p}^{\alpha} 
 \Theta^{\beta \gamma} \bar{p}^{\delta}   \,.
\end{equation}
Notice that $H$ has no linear-in-$S^{ab}$ piece: such terms will come from the geometry \eqref{PBs} alone, which is expected from the universality of linear-in-spin motion \cite{Ha.12}. In Eqs.~\eqref{PBs} and \eqref{HTD}, it is important to appreciate that the phase space coordinates $(x^\alpha,p_\alpha,S^{\alpha\beta})$ can be non-trivially hidden in each term. Taking all these dependencies into account and replacing $F$ by each coordinate in Hamilton's equation gives back Eqs.~\eqref{EE} and \eqref{EElin}, provided that 
$\lambda=\tau/\mu=:\bar{\tau}$. This is a non-trivial calculation that ensures the validity of the whole Hamiltonian framework. As the Hamiltonian \eqref{HTD} is autonomous (independent of $\bar{\tau}$), it is conserved along any solution to Hamilton's equation. Indeed, under Eq.~\eqref{TDSSC}, $H=-\tilde{\mu}^2/2$, which, as expected, must be conserved along $\mathscr{L}$. 

\vspace{1mm}{\it Degeneracy of the system.---} The 14D Poisson manifold $(\mcM,\{,\})$ is degenerate \cite{WiStLu.19}. These degeneracies become obvious when we switch to the variables $(x^\alpha,\pi_\alpha,S^I,D^I)$, where
\begin{equation} \label{rel}
\pi_\alpha :=p_\alpha - \tfrac{1}{2}\omega_{\alpha B C}S^{BC} \,, \,\, S^I := \tfrac{1}{2}\varepsilon^I_{\phantom{I}JK}S^{JK} \,, \,\, D^I:= S^{0I} \,, 
\end{equation}
with $\omega_{aBC}:=g_{bc}(e_B)^b\nabla_a (e_C)^c$ connection 1-forms associated to an arbitrary orthonormal tetrad field $(e_A)^a$ on $(\scE,g_{ab})$, \cite{Wald} (upper case Latin indices denote tetrad components, and $\varepsilon_{IJK}$ is the 3D Levi-Civita symbol with $\varepsilon_{123}=0$). The quantities $(S^I,D^I)$ define two Euclidean 3-vectors $(\vec{S},\vec{D})$ in the tetrad frame, that correspond to the spin and mass dipole of the particle measured by an observer of four-velocity $(e_0)^a$. The Poisson brackets of the new $\mcM$-chart $(x^\alpha,\pi_\alpha,S^I,D^I)$ are found by combining Eqs.~\eqref{PBs} and \eqref{rel} with the Leibniz rule and classical properties of $\omega_{\alpha BC}$ \cite{Wald}. We find $\{x^\alpha,\pi_\beta\} = \delta^\alpha_\beta$ and 
\begin{equation}\label{very}
     \{S^I,S^J\} = \{D^J,D^I\} = \varepsilon^{IJ}_{\phantom{IJ}K} S^K \,, \, 
    \{D^I,S^J\} = \varepsilon^{IJ}_{\phantom{IJ}K} D^K \,.
\end{equation}
The brackets \eqref{very} are a reminder of the symmetry of the underlying Lorentz algebra so(1,3), inherited from the orthonormal tetrad. Crucially, the spin-curvature coupling in the last bracket of \eqref{PBs} is now absorbed into the new coordinate $\pi_\alpha$, making the pair $(x^\alpha,\pi_\alpha)$ geometrically independent of the spin sector $(S^I,D^I)$. The degeneracy of the Poisson structure \eqref{very} can now be easily established. Consider the Poisson matrix $\Lambda$ expressed in the new coordinates:
\begin{equation} \label{sympN}
\Lambda=\text{diag}(\mathbb{J}_8,\mathfrak{S}) \,, \,\, \mathbb{J}_8=\left( \begin{array}{cc}
  0 & \mathbb{I}_4 \\
-\mathbb{I}_4 &0 
\end{array} \right) \,, \,\, 
\fS=\left( \begin{array}{cc}
  \mathcal{S} & \mathcal{D} \\
\mathcal{D} & -\mathcal{S} 
\end{array} \right) \,,
\end{equation}
where $\mathbb{I}_4$ is the $4\times 4$ identity matrix, so that $\mathbb{J}_8$ is the canonical $8\times 8$ Poisson matrix, while $\mathfrak{S}$ is a $6\times 6$ antisymmetric matrix constructed from SO(3) matrices associated with $S^I$ and $D^I$:
\begin{equation}
\mathcal{S}=\left( \begin{array}{ccc}
  0 & S^3 & -S^2 \\
  -S^3 & 0 & S^1 \\
  S^2 & -S^1 & 0 \\
\end{array} \right)
\,, \,\,
\mathcal{D}=\left( \begin{array}{ccc}
  0 & D^3 & -D^2 \\
  -D^3 & 0 & D^1 \\
  D^2 & -D^1 & 0 \\
\end{array} \right) \,.
\end{equation}
Direct inspection reveals that $\text{rank}(\Lambda)=12$: The Poisson structure $\Lambda$ is degenerate. According to Poisson systems theory, this degeneracy is associated with the existence of $\text{dim}(\mcM)-\text{rank}(\Lambda)=2$ \textit{Casimir invariants} \cite{Vaisman.12,Derigl.22}. These invariants, denoted $\mathcal{C}_\circ,\mathcal{C}_\star$, are obtained by direct calculation of the null space of $\Lambda$, wherein their gradients form a basis. One finds $\mathcal{C}_\circ=
\vec{S}\cdot\vec{S}-\vec{D}\cdot\vec{D}$ and $\mathcal{C}_{\star} =
\vec{S}\cdot\vec{D}$, using Euclidean 3-vectors notations. 
Such degeneracies are unavoidable for relativistic spinning objects: They are tied to local Lorentz invariance of general relativity. It is easily checked that the pair $(\mathcal{C}_{\circ},\mathcal{C}_{\star})$ coincides physically with the spin norms $(S_\circ^2,S_\star^2)$ \cite{WiStLu.19}. Therefore, in our framework, the constancy of $S_\circ$ is a consequence of phase space geometry only (independent of the Hamiltonian).

\vspace{1mm}{\it Symplectic formulation.---} The classical notion of Liouville-Arnold integrability requires a symplectic (non-degenerate) formulation of the dynamics. Weinstein splitting theorem's for Poisson manifolds \cite{Vaisman.12} shows that the 14D Poisson manifold $(\mcM,\Lambda)$ is foliated by 12D, symplectic sub-manifolds known as \textit{symplectic leaves}, corresponding to the level sets of the Casimirs $(\mcC_1,\mcC_2)$. Let $\mcN$ be such a leaf, where the Casimirs have physical value $(S_\circ^2,S_\star^2)$. On $\mcN$, there exists a natural non-degenerate Poisson structure inherited from that of $\mcM$. Symplecticity means that it is possible, at least locally, to endow $\mcN$ with 12 \textit{canonical coordinates}, i.e., 6 pairs $(q^i,\pi_{q_i})_{i\in\{1,\ldots,6\}}$ for which the brackets are $\{q^i,\pi_{q_j}\}=\delta^i_j$. These coordinates can be built explicitly \cite{Ra.PapI.24} but are not required for our present purposes. The Hamiltonian on $\mcN$ is obtained by inserting the new coordinates \eqref{rel} in the Hamiltonian \eqref{HTD} on $\mcM$. One finds
\begin{equation}\label{Htot}
    H_\mcN =\frac{1}{2}g^{\alpha\beta}\pi_\alpha\pi_\beta + \frac{1}{2} g^{\alpha\beta}\pi_\alpha\omega_{\beta CD}S^{CD} + Q_{ABCD} S^{AB} S^{CD} \,,
\end{equation}
where $Q_{ABCD} := \tfrac{1}{8} \omega^{\alpha}_{\phantom{\alpha} AB} \omega_{\alpha CD} + R_{EABD} \bar{\pi}^E \bar{\pi}_C + \tfrac{\kappa}{2} R_{EADF} \bar{\pi}^E \bar{\pi}^F \eta_{BC}$ depends only on $(x^\alpha,\pi_\alpha)$ and should be set to zero to recover the linear-in-spin dynamics studied in \cite{Ra.PapI.24,Ra.Iso.PapII.24}. We note that the quantities $\bar{\pi}^A:=\eta^{AB}(e_B)^\alpha \pi_\alpha/\mu$ depend on the coordinates $x^\alpha$ (via $(e_B)^\alpha$ and $\mu$) and $\pi_\alpha$ (explicitly and in $\mu$), while $S^{AB}$ are simple functions of $(S^I,D^I)$ hidden through $S^{0I}=D^I$ and $S^{IJ}=\varepsilon^{IJ}_{\phantom{IJ}K}S^K$, cf. Eq.~\eqref{rel}. The Hamiltonian \eqref{Htot} on the 12D leaves $\mcN$, along with the 12 canonical coordinates given in \cite{Ra.PapI.24} and canonical Poisson brackets, define a 12D symplectic Hamiltonian system. This system, however, is not physically satisfactory: it contains unphysical trajectories and hosts spurious chaotic regions unrelated to physical effects. This is a crucial point of our analysis, which we now detail.

\vspace{1mm}{\it Hamiltonian status of the SSC.---} The phase space $\mcN$ is 12-dimensional, even though the physics at hand, with the three independent SSC constraints \eqref{TDSSC} and the normalization condition $u_\alpha u^\alpha=-1$, present only $14-3-1=10$ independent unknowns. In addition, $\mcN$ is filled with trajectories that are non-physical, namely those that do not satisfy the SSC \eqref{TDSSC}. These issues come from a subtlety of the MPTD system scarcely noted and worth discussing. Equations \eqref{EE}-\eqref{EElin} are a consequence of the SSC \eqref{TDSSC}, but does not imply it back: Eqs.~\eqref{EE}-\eqref{EElin} only imply $\nabla_uC^a=0$. Parallel transport along $\scL$ being linear, if $C^a=0$ at some point on $\scL$, then $\nabla_uC^a=0$ implies $C^a=0$ at each point of $\scL$. In other words, when working at the level of the evolution equations along $\scL$ with some initial conditions, this issue is harmless. However, in a symplectic framework, there is no way to ``choose'' such initial condition. The phase space $\mcM$ and its leaves $\mcN$ are swarmed with trajectories that satisfy Hamilton's equations \eqref{EElin} but not the SSC \eqref{TDSSC}, making them physically irrelevant. \vspace{2mm}

What one must do is \textit{restrict} the analysis to the sub-manifold $\mcT\subset\mcM$ where the $p_\alpha S^{\alpha\beta}=0$ holds \textit{identically}. 
This sub-manifold has two important properties. First, it has co-dimension 3 in $\mcM$, as \eqref{TDSSC} only has 3 linearly independent components. Second, $\mcT$ is \textit{invariant} under the flow of the Hamiltonian \eqref{HTD}, as Eqs.~\eqref{TDSSC} and \eqref{PBs} readily imply\footnote{This equation can be obtained by a direct combination of Eqs.~\eqref{PBs} and \eqref{HTD}, or from the fact that Eqs.~\eqref{EE} and \eqref{EElin} imply $\nabla_u C^a=0$, which is a rewriting of \eqref{invP} given that $\{C^\alpha,H\}=\ud C^\alpha/\ud\bar{\tau}$.}
\begin{equation} \label{invP}
    \{C^\alpha,H\} = - \Gamma^\alpha_{\beta\gamma}p^\beta C^\gamma \,, 
\end{equation}
up to cubic-in-$S^{ab}$ corrections. Therefore, any trajectory that starts on $\mcT$ will stay on it at the perturbative ordered considered.  
Both issues mentioned above (too many degrees of freedom and non-physical trajectories) are resolved by restricting our analysis to the sub-manifold $\mcT$. More precisely, we look at the intersections $\mcP:=\mcT\cap\mcN$, where the SSC holds identically \textit{and} Casimir degeneracies have been lifted (this is the notion of cosymplectic sub-manifolds \cite{Zambon.11,Vaisman.12}). Writing the SSC \eqref{TDSSC} in terms of the variables on $\mcN$ leads to
$\pi_I D^I = N_{JK} S^J S^K$ and $\pi_0 D^I =\varepsilon^I_{\phantom{I}JK} \pi^J S^K+ M^I_{\phantom{I}JK} S^J S^K$, 
where $N_{JK},M^I_{\phantom{I}JK}$ only depend on $(x^\alpha,\pi_\alpha)$ via $\omega_{\alpha BC}$ and $\pi_A$. These equations readily imply $\vec{D}\cdot\vec{S}=0$ to quadratic order in spin. Geometrically, this means that $\mcT$ only intersects symplectic leaves where $\mcC_2=0$, and the dimension of $\mcP$ depends on the leaf $\mcN$: $\text{dim}(\mcP)=10$ if $\mcC_2=0$ and $\text{dim}(\mcP)=9$ otherwise. We focus on the former case, as it is the only one of physical interest, and let $\mcN$ be any leaf where $\mcC_\star=0$ from now on. It follows from the above discussion that only two out of the four SSC constraints are sufficient to define $\mcP$. In theory, any pair can be used. In practice, we use
\begin{equation}\label{D0D1}
    \begin{split}
        C^0 &:= -\pi_I D^I + N_{IJ}S^IS^J = 0\,, \\
        C^1 &:= \pi_0 D^1 - \varepsilon^1_{\phantom{1}IJ} \pi^I S^J- M^1_{\phantom{1}IJ}S^IS^J = 0\,,
        \end{split}
\end{equation}
so that they coincide with the first two tetrad components of the SSC \eqref{TDSSC}, namely, $C^B:=p_A S^{AB}$. 

\vspace{1mm}{\it Symplectic structure on $\mcP$.---} Following the classical theory of constrained Hamiltonian systems \cite{Derigl.22}, on the 10D sub-manifold $\mcP$ where \eqref{D0D1} holds, there exists a symplectic structure inherited from that on $\mcN$, provided that $\{C^0,C^1\}|_\mcP \neq 0$. This is readily verified by combining Eqs.~\eqref{TDSSC} and \eqref{PBs} to find 
\begin{equation} \label{MAB}   
\{ C^{\alpha}, C^{\beta} \} =- \mu^2 S^{\alpha \beta} - 2 p^{[\alpha}
 C^{\beta ]} - 2 S^{\gamma [\alpha}
\Gamma^{ \beta]}_{\gamma \delta} C^{\delta}\,. 
\end{equation}
Projecting onto the tetrad and using \eqref{D0D1} gives, on $\mcP$ only, $\{C^0,C^1\}=-\mu^2 D^1\neq 0$. Consequently, $(\mcP,\Lambda_\mcP)$ is a well-defined 10D symplectic manifold, with $\Lambda_\mcP$ the restriction of $\Lambda$ to $\mcP$. The Poisson bracket on $\mcP$, hereafter denoted $\{,\}^\mcP$, are obtained by pulling back the symplectic structure $\Lambda$ on $\mcN$ via the embedding $\mcP \hookrightarrow \mcN$ \cite{Zambon.11,Burs.13,Derigl.22}. One finds 
\begin{equation} \label{Dirac}
\{ F, G \}^\mcP := \{ F, G \} - \frac{\{ F, C^0 \} \{ G, C^1 \}- \{ F, C^1 \} \{ G, C^0 \}}{\{ C^0, C^1 \}} \,, 
\end{equation}
in which the right-hand side is to be computed with the brackets $\{,\}$ on $\mcN$ first, and then evaluated on $\mcP$, i.e., simplified with Eqs.~\eqref{D0D1}. The 10D phase space $\mcP$, the brackets \eqref{Dirac} and the Hamiltonian $H_\mcP$ (obtained from \eqref{Htot} expressed in terms of 10 well-chosen coordinates on $\mcP$), define a Hamiltonian system $(H_\mcP,\Lambda_\mcP,\mcP)$, which (i) generates system \eqref{EE}--\eqref{EElin} and (ii) has the Tulczyjew SSC \eqref{TDSSC} built-in. It is the one in which we now prove our integrability results. 

\vspace{1mm}{\it Conserved quantities.---} Our definition of integrability is the classical one --Liouville-Arnold \cite{Liou1855,Arn}--, precisely following \cite{HiFl.08} for the relativistic context. In particular, the system $(\mcP,\{,\}^\mcP,H_\mcP)$ will be \textit{integrable} if there exists five first integrals $(\mathcal{I}_1,\ldots,\mathcal{I}_5)$ that are linearly independent and in pairwise involution.\footnote{
As always in a perturbative setup, the notion of integrability that we use is that of pairwise commuting first integrals up to some order in a small parameter, as used in \cite{Card.al.18,Tanay.al.21} and described in \cite{FaMa.06}. %A well-chosen coordinate transformation can always make this approximate notion exact \cite{Fejozprivate}.   
} In general relativity, many spacetime of interest possess symmetries associated with Killing fields of various rank. These Killing fields are associated with conserved quantities of the MPTD system \eqref{EE} and have been extensively studied in the past \cite{Di.64,Rudiger.I.81,Rudiger.II.83,Ha.15,ComDru.22,HaDw.23}. We shall build upon these works to find first integrals in our Hamiltonian framework. In particular, let $k^a$ be a Killing vector and $Y^{ab}$ a Killing-Yano tensor of $g_{ab}$. Consider the following quantities:
%
%\begin{equation}\label{Killinvs}\begin{split}
%    \Xi &:=p_a k^a + \tfrac{1}{2} S^{ab}\nabla_a k_b  \,, \quad
%    \fK := \tfrac{1}{4}\varepsilon_{abcd} Y^{ab} S^{cd} \,, \\
%    \fQ &:= Y^{a}_{~c} Y^{bc} p_a p_b +4  \varepsilon_{ade[b} Y^{e}_{~c]} \xi^a S^{db} p^c + M_{abcd}S^{ab}S^{cd}, 
%    \end{split}\end{equation} 
%
\begin{subequations}\label{Killinvs}
    \begin{align}
    \Xi &:=p_a k^a + \tfrac{1}{2} S^{ab}\nabla_a k_b  \,, \\
    \fK &:= \tfrac{1}{4}\varepsilon_{abcd} Y^{ab} S^{cd} \,, \\
    \fQ &:= K^{ab} p_a p_b +4  \varepsilon_{ade[b} Y^{e}_{~c]} \xi^a S^{db} p^c + M_{abcd}S^{ab}S^{cd} \,,
    \end{align}
\end{subequations} 
where $K^{ab}:=Y^{a}_{~c}$ is a symmetric Killing-Stäckel tensor and $M_{abcd}:=g_{ac}(\xi_b \xi_d-\tfrac{1}{2}g_{bd}\xi^e\xi_e)-\tfrac{1}{2}Y_a^{\phantom{a}e}(Y_c^{\phantom{a}f}R_{ebfd} +\tfrac{1}{2} Y_e^{\phantom{e}f}R_{fbcd})$, with $\xi^a$ defined\footnote{Such that $\nabla_{a}Y_{bc}:=\xi^d\varepsilon_{dabc}$ and $\xi^a=(\partial_t)^a$ in Kerr.} as $\xi^a:=-\frac{1}{6} \varepsilon^{abcd}\nabla_bY_{cd}$. The quantities $(\Xi,\fK,\fQ)$ have been shown to be conserved along $\scL$ under different assumptions regarding (i) the spin order considered, (ii) the SSC, (ii) the background metric and (iv) the nature of the secondary body. These conditions are gathered from \cite{Di.64,Rudiger.I.81,Rudiger.II.83,Ha.12,ComDru.22,ComDruVin.23} and summarized in Table \ref{table}. As $(\Xi,\fK,\fQ)$ have been constructed at the level of the MPTD system in the literature, they are natural candidates for first integrals of the Hamiltonian system $(H_\mcP,\{,\}^\mcP,\mcP)$.

\begin{table}[!ht]
	\begin{tabular}{c|c|c|c|c}
		\toprule
            \textbf{Constant} \,\, &  \,\,\,\, \textbf{Order}\,\, \,\, & \,\,\textbf{SSC}   \,\,&\,\,  \textbf{Metric}   \,\, &\,\, \textbf{Body}  \,\, \\ 
        $\Xi$                     &  all  & any   &  any    & any       \\ 
		$\fK,\fQ$                 &  linear  & TD   &  Ricci-flat    & any                    \\ 
  	$\fK,\fQ$                 &  quadratic  & TD   &  Kerr    & $\kappa=1$                   \\ 
	\end{tabular} \caption{Conditions for the conservation of the constants of motion.}
    \label{table}
\end{table}

\vspace{1mm}{\it Integrability in Kerr.---} We now apply the framework to the Kerr spacetime $(\scE,g_{ab})$, where $\scE$ is covered by Boyer-Lindquist coordinates $x^\alpha=(t,r,\theta,\phi)$ and $g_{ab}$ is the Kerr metric with mass $M \geq 0$ and spin parameter $a\in[0,M]$. The tetrad $(e_A)^a$ that we chose is the Carter tetrad, admitting 10 independent non-vanishing connection 1-forms $\omega_{aBC}$ \cite{ViKuStHi.16,Ra.PapI.24}. The two Killing vectors $(\partial_t)^a$ and $(\partial_\phi)^a$ generate the invariants $E := -\pi_t$ and $L_z := \pi_\phi$, interpreted as the object's total energy and angular momentum component, measured at spatial infinity \cite{GourgoulhonBH}. The Killing-Yano tensor $Y^{ab}:=2r (e_2)^{[a}(e_3)^{b]}-2a\cos\theta (e_0)^{[a}(e_1)^{b]}$ \cite{Floyd.73} generates two additional invariants \eqref{Killinvs}, namely $\fK := r \, D^1 + a\cos\theta \, S^1$, and $\fQ$ which is too long to be displayed here but can be found explicitly in the Mathematica Notebook \cite{MMAPRL}. 

Explicit calculations of the relevant Poisson brackets in a Kerr background reveal that: (i) $\{E,H_\mcP \}^\mcP=\{L_z,H_\mcP \}^\mcP =0$ at linear and quadratic order in spin; (ii) $\{\fK,H_\mcP\}^\mcP=\{\fQ,H_\mcP\}^\mcP=0$ at linear order in spin (i.e., when setting 
$Q_{ABCD}=0$ in \eqref{Htot} and $M_{abcd}=0$ in \eqref{Killinvs}); and (iii) 
$\{\fK,H_\mcP\}^\mcP$ and 
$\{\fQ,H_\mcP\}^\mcP$ are both proportional to $(\kappa-1)$ at quadratic order in spin. Along with $H_\mcP=-\tilde{\mu}^2/2$ itself, we have five first integrals $(H_\mcP,E,L_z,\fK,\fQ)$ at linear order in spin for any compact object, and at quadratic order in spin for objects with $\kappa=1$. 

The five first integrals are easily shown to be linearly independent. Integrability will thus follow from their pairwise involution. To verify this, there are $\binom{5}{2}=10$ independent $\{,\}^\mcP$-brackets to check. Nine of them involve at least one instance of either $H_\mcP,E$ or $L_z$. These brackets vanish by general arguments about the Hamiltonian and Killing vectors, cf.~\cite{Ra.PapI.24}, leaving only \textit{one} non-trivial bracket to compute, namely $\{\fK,\fQ\}^\mcP$. We compute it explicitly using Eqs.~\eqref{Dirac} and then applying \eqref{D0D1}. This computation is tractable by hand to linear order in spin in a Schwarzschild background. The quadratic-in-spin case and/or Kerr cases are performed in a dedicated Mathematica Notebook \cite{MMAPRL}. It reveals that, to quadratic order in spin,
\begin{equation} \label{kq}
    \{\fK,\fQ\}^\mcP \propto (\kappa-1) \,, 
\end{equation}
up to cubic-in-spin corrections, even though $\{\fK,\fQ\} \neq 0$ in general, even to linear order in spin. This contrast reflects the importance of a rigorous Hamiltonian treatment of the SSC, i.e., using the $\mcP$-brackets. The result \eqref{kq} alone concludes the proof of linear-in-spin integrability in Kerr for any compact object, and of quadratic-in-spin integrability in Kerr for a compact object with $\kappa=1$. 

For $\kappa\neq1$, quadratic-in-spin integrability is compromised: There are strictly less than $5$ first integrals (we found 4 in Schwarzschild\footnote{Remarkably, in Schwarzschild, $\fQ$ is still conserved for $\kappa\neq 1$ at quadratic-in-$S^{ab}$ order, because it coincides with the square of the total angular momentum $L^2$ built from the Killing vectors of the SO(3) symmetry, even though the definitions of $L^2$ and $\fQ$ are completely unrelated in general.} and 3 in Kerr). One may attribute this feature to objects with $\kappa\neq 1$ being, somehow, ``less symmetric'' than those with $\kappa=1$, the latter including highly ``symmetric'' isolated black holes. 

But why $\kappa=1$ in particular? While still puzzling at a fundamental level, we looked precisely at intermediary steps of the calculations and observed that some quadratic-in-spin terms arising from the dipole-dipole interactions (sourced by the second term on the right-hand side of Eq.~\eqref{EElin}) were exactly cancelled by the quadrupolar contributions (like the third term there) if and only if $\kappa=1$. This is somewhat expected: black holes' are unique in that their multipoles satisfy non-trivial, exact relations. Thus, there can be combinations of multipole contributions of different orders that have a chance to cancel each other. To illustrate, an isolated Kerr black hole's monopolar, dipolar and quadrupolar moments are given by $M,a$ and $Q=-a^2M$ \cite{Thorne.80}. Thus, schematically, one has that $[\text{quadrupole}]+[\text{monopole}]\times[\text{dipole}]^2 = 0$. For families of compact objects with multipole moments depending on their equation of state such relations between multipoles are not exact as for black holes but only approximate and microphysics dependent \cite{YaYu.13,YaYu.16}.

\vspace{1mm}{\it Applications to waveform modelling.---} Perhaps the most crucial implication of integrability is that, to describe the linear-in-spin motion of a body orbiting a Kerr black hole, there exists 5 pairs of action-angle variables $(\vartheta^i,\mathcal{J}_i)_{i=1\ldots 5}$ on $\mcP$ \cite{Arn}. The angles $\vartheta^i$ are cyclic (the Hamiltonian does not depend on them) and the actions $\mathcal{J}_i$ are first integrals, in a 1-to-1 correspondence with the set $(H_\mcP,E,L_z,\fQ,\fK)$. This has a strong potential to improve current gravitational templates waveform generation schemes \cite{HiFl.08,PoWa.21} developed for compact binary systems with asymmetric mass ratio, a prime targets of the LISA mission \cite{LISAWHITE.23}. 

Currently, this scheme relies heavily on the integrability of Kerr geodesics, implying the existence of four pairs of action-angle variables \cite{Schm.02,HiFl.08}. While the effect of the secondary's spin, treated as a first post-adiabatic order \textit{perturbation}, slowly evolves the geodesic constants of motion, our result suggests that it is possible to include these spin effects in the five action-angle variables $(\vartheta^i,\mathcal{J}_i)$, i.e., and use the spinning trajectory itself as the integrable basis, on which to add the conservative self-force. These linear-in-spin corrections would be automatically included into the constants of motion everywhere in the parameter space, consistent with the inspiral evolution schemes in \cite{MaPoWa.22}. In addition, analytically solving the spinning trajectories by quadrature, as allowed by Arnold-Liouville integrability, would bypass the need for pre-computed trajectories, for which current state-of-the-art methods rely on numerical integration \cite{DruHug.I.22,DruHug.II.22}. Given the size of the parameter space for generic orbits, this possibility represents a significant reduction in the computational cost of waveform generation at the accuracy required by LISA \cite{LISAWHITE.23}. 

Similarly, the derivation of flux-balance laws \cite{IsoAl.19} for the integrals of motion (particularly with the recent improved understanding of hidden Kerr symmetries \cite{GrMo.23}) and of the first law of binary mechanics in Kerr \cite{Is.al.14,Le.15}, both relying entirely on the integrability of Kerr geodesics, can now be extended to the spinning, extended-body case. These laws are critical to extract conservative and dissipative effects of the gravitational self-force \cite{Po.al.20} and benchmark the predictions of different approximation schemes \cite{Le2.14}. Still regarding the self-force, our covariant Hamiltonian framework could now be used as a basis to add its conservative piece by adapting the methods of \cite{IsoAl.19,BlFl2.23}, and the dissipative part by using Galley's theory of non-conservative systems \cite{Galley.13}, and investigate the integrability \cite{Nasipak.22,BlFl.23}. 

\vspace{1mm}{\it Future prospects.---} Our Hamiltonian framework offers natural, straightforward extensions, from spin effects at higher multipolar orders \cite{Ma.15} to tidal quadrupolar models \cite{StPu.10,RaLe.20}, all in any background spacetimes. This will contribute to the ongoing effort in elucidating the links between integrability and local symmetries at quadrupolar and/or quadratic-in-spin orders \cite{Frolov.17,Tanay.al.21,ComDruVin.23,HaDw.23}. The overlap of all these works would be worth investigating to reveal, perhaps, a bigger picture regarding integrability of compact objects binaries. In addition, Hamiltonian systems are tailor-made for structure-preserving numerical schemes, such as symplectic integrators \cite{Zho.al.10,Sey.Lu.12,Wu.Kerr.21}. As these can substantively increase computational efficiency and long-term stability of numerical solutions, combining our framework with these numerical schemes gives a promising opportunity to investigate the quadratic-in-spin dynamics. 

Our result on integrability opens the possibility to extend well-known Kerr geodesic features to the spinning-case. Integrability implies the existence of gauge-independent frequencies $\Omega_i:=\partial H/\partial \mathcal{J}_i$, allowing one to extend the current understanding of resonant motion of Kerr geodesics \cite{BrGeHi.15} to spinning trajectories, now characterized by a three-fold resonances involving a spin-precession frequency in addition to azimuthal and radial components \cite{MukTri.20}. Whether quadratic-in-spin effects enhance or soften transient resonances is important for LISA, as virtually all its asymmetric binary sources will pass at least through one once \cite{RuHu.14}. In addition, spin-induced quadrupoles will be invaluable to probe the state of matter in compact stars using the next generation of ground-based interferometers \cite{ET.al.20,CE.22}. Our framework gives a simple and powerful framework to further investigate these effects.

\vspace{1mm}{\it Acknowledgments.---}I am indebted to A.~Druart and S.~Isoyama for independent verifications of the results. I also thank F.~Blanco, L.~Drummond, A.~Harte, S.~Hughes,  T.~Hinderer, A.~Le Tiec, J.~Mathews, L.~Stein, S.~Tanay and K.~Van Aelst for illuminating discussions and valuable feedback.

%\bibliography{ListeRef.bib}

\begin{thebibliography}{86}%
\makeatletter
\providecommand \@ifxundefined [1]{%
 \@ifx{#1\undefined}
}%
\providecommand \@ifnum [1]{%
 \ifnum #1\expandafter \@firstoftwo
 \else \expandafter \@secondoftwo
 \fi
}%
\providecommand \@ifx [1]{%
 \ifx #1\expandafter \@firstoftwo
 \else \expandafter \@secondoftwo
 \fi
}%
\providecommand \natexlab [1]{#1}%
\providecommand \enquote  [1]{``#1''}%
\providecommand \bibnamefont  [1]{#1}%
\providecommand \bibfnamefont [1]{#1}%
\providecommand \citenamefont [1]{#1}%
\providecommand \href@noop [0]{\@secondoftwo}%
\providecommand \href [0]{\begingroup \@sanitize@url \@href}%
\providecommand \@href[1]{\@@startlink{#1}\@@href}%
\providecommand \@@href[1]{\endgroup#1\@@endlink}%
\providecommand \@sanitize@url [0]{\catcode `\\12\catcode `\$12\catcode `\&12\catcode `\#12\catcode `\^12\catcode `\_12\catcode `\%12\relax}%
\providecommand \@@startlink[1]{}%
\providecommand \@@endlink[0]{}%
\providecommand \url  [0]{\begingroup\@sanitize@url \@url }%
\providecommand \@url [1]{\endgroup\@href {#1}{\urlprefix }}%
\providecommand \urlprefix  [0]{URL }%
\providecommand \Eprint [0]{\href }%
\providecommand \doibase [0]{http://dx.doi.org/}%
\providecommand \selectlanguage [0]{\@gobble}%
\providecommand \bibinfo  [0]{\@secondoftwo}%
\providecommand \bibfield  [0]{\@secondoftwo}%
\providecommand \translation [1]{[#1]}%
\providecommand \BibitemOpen [0]{}%
\providecommand \bibitemStop [0]{}%
\providecommand \bibitemNoStop [0]{.\EOS\space}%
\providecommand \EOS [0]{\spacefactor3000\relax}%
\providecommand \BibitemShut  [1]{\csname bibitem#1\endcsname}%
\let\auto@bib@innerbib\@empty
%</preamble>
\bibitem [{\citenamefont {Mathisson}(1937)}]{Ma.37}%
  \BibitemOpen
  \bibfield  {author} {\bibinfo {author} {\bibfnamefont {M.}~\bibnamefont {Mathisson}},\ }\href@noop {} {\bibfield  {journal} {\bibinfo  {journal} {Acta Phys. Polon.}\ }\textbf {\bibinfo {volume} {6}},\ \bibinfo {pages} {163} (\bibinfo {year} {1937})}\BibitemShut {NoStop}%
\bibitem [{\citenamefont {Dixon}(1964)}]{Di.64}%
  \BibitemOpen
  \bibfield  {author} {\bibinfo {author} {\bibfnamefont {W.~G.}\ \bibnamefont {Dixon}},\ }\href@noop {} {\bibfield  {journal} {\bibinfo  {journal} {Il Nuovo Cimento}\ }\textbf {\bibinfo {volume} {34}},\ \bibinfo {pages} {317} (\bibinfo {year} {1964})}\BibitemShut {NoStop}%
\bibitem [{\citenamefont {Harte}(2015)}]{Ha.15}%
  \BibitemOpen
  \bibfield  {author} {\bibinfo {author} {\bibfnamefont {A.~I.}\ \bibnamefont {Harte}},\ }\href@noop {} {\bibfield  {journal} {\bibinfo  {journal} {Fund. Theor. Phys.}\ }\textbf {\bibinfo {volume} {179}},\ \bibinfo {pages} {327} (\bibinfo {year} {2015})}\BibitemShut {NoStop}%
\bibitem [{\citenamefont {Hartl}(2003)}]{Hartl.03}%
  \BibitemOpen
  \bibfield  {author} {\bibinfo {author} {\bibfnamefont {M.~D.}\ \bibnamefont {Hartl}},\ }\href@noop {} {\bibfield  {journal} {\bibinfo  {journal} {Phys. Rev. D}\ }\textbf {\bibinfo {volume} {67}},\ \bibinfo {pages} {104023} (\bibinfo {year} {2003})}\BibitemShut {NoStop}%
\bibitem [{\citenamefont {Kao}\ and\ \citenamefont {Cho}(2005)}]{KaoCho.05}%
  \BibitemOpen
  \bibfield  {author} {\bibinfo {author} {\bibfnamefont {J.~K.}\ \bibnamefont {Kao}}\ and\ \bibinfo {author} {\bibfnamefont {H.~T.}~\bibnamefont {Cho}},\ }\href@noop {} {\bibfield  {journal} {\bibinfo  {journal} {Phys. Lett. A}\ }\textbf {\bibinfo {volume} {336}},\ \bibinfo {pages} {159} (\bibinfo {year} {2005})}\BibitemShut {NoStop}%  
\bibitem [{\citenamefont {Han}(2008)}]{Han.08}%
  \BibitemOpen
  \bibfield  {author} {\bibinfo {author} {\bibfnamefont {W.}\ \bibnamefont {Han}},\ }\href@noop {} {\bibfield  {journal} {\bibinfo  {journal} {General Relativity and Gravitation}\ }\textbf {\bibinfo {volume} {40}},\ \bibinfo {pages} {1831} (\bibinfo {year} {2008})}\BibitemShut {NoStop}%
\bibitem [{\citenamefont {Hartl2}(2003)}]{Hartl2.03}%
  \BibitemOpen
  \bibfield  {author} {\bibinfo {author} {\bibfnamefont {M.~D.}\ \bibnamefont {Hartl}},\ }\href@noop {} {\bibfield  {journal} {\bibinfo  {journal} {Phys. Rev. D}\ }\textbf {\bibinfo {volume} {67}},\ \bibinfo {pages} {024005} (\bibinfo {year} {2003})}\BibitemShut {NoStop}%
\bibitem [{\citenamefont {Suzuki}\ and\ \citenamefont {Maeda}(1999)}]{SuMa.99}%
  \BibitemOpen
  \bibfield  {author} {\bibinfo {author} {\bibfnamefont {S.}\ \bibnamefont {Suzuki}}\ and\ \bibinfo {author} {\bibfnamefont {K.}~\bibnamefont {Maeda}},\ }\href@noop {} {\bibfield  {journal} {\bibinfo  {journal} {Phys. Rev. D}\ }\textbf {\bibinfo {volume} {61}},\ \bibinfo {pages} {024005} (\bibinfo {year} {1999})}\BibitemShut {NoStop}%  
\bibitem [{\citenamefont {Suzuki}\ and\ \citenamefont {Maeda}(1997)}]{SuMa.97}%
  \BibitemOpen
  \bibfield  {author} {\bibinfo {author} {\bibfnamefont {S.}\ \bibnamefont {Suzuki}}\ and\ \bibinfo {author} {\bibfnamefont {K.}~\bibnamefont {Maeda}},\ }\href@noop {} {\bibfield  {journal} {\bibinfo  {journal} {Phys. Rev. D}\ }\textbf {\bibinfo {volume} {55}},\ \bibinfo {pages} {4848} (\bibinfo {year} {1997})}\BibitemShut {NoStop}%  
\bibitem [{\citenamefont {Yagi}\ and\ \citenamefont {Yunes}(2013)}]{YaYu.13}%
  \BibitemOpen
  \bibfield  {author} {\bibinfo {author} {\bibfnamefont {K.}\ \bibnamefont {Yagi}}\ and\ \bibinfo {author} {\bibfnamefont {N.}~\bibnamefont {Yunes}},\ }\href@noop {} {\bibfield  {journal} {\bibinfo  {journal} {Science}\ }\textbf {\bibinfo {volume} {341}},\ \bibinfo {pages} {365} (\bibinfo {year} {2013})}\BibitemShut {NoStop}%  
\bibitem [{\citenamefont {Yagi}\ and\ \citenamefont {Yunes}(2016)}]{YaYu.16}%
  \BibitemOpen
  \bibfield  {author} {\bibinfo {author} {\bibfnamefont {K.}\ \bibnamefont {Yagi}}\ and\ \bibinfo {author} {\bibfnamefont {N.}~\bibnamefont {Yunes}},\ }\href@noop {} {\bibfield  {journal} {\bibinfo  {journal} {Classical and Quantum Gravity}\ }\textbf {\bibinfo {volume} {33}},\ \bibinfo {pages} {095005} (\bibinfo {year} {2016})}\BibitemShut {NoStop}%  
\bibitem [{\citenamefont {Mathews}\ \emph {et~al.}(2022)\citenamefont {Mathews}, \citenamefont {Pound},\ and\ \citenamefont {Wardell}}]{MaPoWa.22}%
  \BibitemOpen
  \bibfield  {author} {\bibinfo {author} {\bibfnamefont {J.}~\bibnamefont {Mathews}}, \bibinfo {author} {\bibfnamefont {A.}~\bibnamefont {Pound}}, \ and\ \bibinfo {author} {\bibfnamefont {B.}~\bibnamefont {Wardell}},\ }\href@noop {} {\bibfield  {journal} {\bibinfo  {journal} {Phys. Rev. D}\ }\textbf {\bibinfo {volume} {105}},\ \bibinfo {pages} {084031} (\bibinfo {year} {2022})}\BibitemShut {NoStop}%
\bibitem [{\citenamefont {Einstein}(1914)}]{Einstein.14}%
  \BibitemOpen
  \bibfield  {author} {\bibinfo {author} {\bibfnamefont {A.}~\bibnamefont {Einstein}},\ }\href@noop {} {\bibfield  {journal} {\bibinfo  {journal} {Sitzungsber. Preuss. Akad. Wiss. Berlin (Math. Phys.)}\ }\textbf {\bibinfo {volume} {1914}},\ \bibinfo {pages} {1030} (\bibinfo {year} {1914})}\BibitemShut {NoStop}%
\bibitem [{\citenamefont {Harte}\ and\ \citenamefont {Dwyer}(2023)}]{HaDw.23}%
  \BibitemOpen
  \bibfield  {author} {\bibinfo {author} {\bibfnamefont {A.~I.}\ \bibnamefont {Harte}}\ and\ \bibinfo {author} {\bibfnamefont {D.}~\bibnamefont {Dwyer}},\ }\href@noop {} {\bibfield  {journal} {\bibinfo  {journal} {Phys. Rev. D}\ }\textbf {\bibinfo {volume} {108}},\ \bibinfo {pages} {124005} (\bibinfo {year} {2023})}\BibitemShut {NoStop}%
\bibitem [{\citenamefont {Damour}\ and\ \citenamefont {Nagar}(2016)}]{DaNa}%
  \BibitemOpen
  \bibfield  {author} {\bibinfo {author} {\bibfnamefont {T.}~\bibnamefont {Damour}}\ and\ \bibinfo {author} {\bibfnamefont {A.}~\bibnamefont {Nagar}},\ }\href@noop {} {\emph {\bibinfo {title} {The effective-one-body approach to the general relativistic two body problem}}},\ \bibinfo {series} {Lecture Notes in Physics}, Vol.\ \bibinfo {volume} {905}\ (\bibinfo  {publisher} {Springer},\ \bibinfo {address} {New York},\ \bibinfo {year} {2016})\BibitemShut {NoStop}%
\bibitem [{\citenamefont {Fujita}\ \emph {et~al.}(2017)\citenamefont {Fujita}, \citenamefont {Isoyama}, \citenamefont {Le~Tiec}, \citenamefont {Nakano}, \citenamefont {Sago},\ and\ \citenamefont {Tanaka}}]{Fu.al.17}%
  \BibitemOpen
  \bibfield  {author} {\bibinfo {author} {\bibfnamefont {R.}~\bibnamefont {Fujita}}, \bibinfo {author} {\bibfnamefont {S.}~\bibnamefont {Isoyama}}, \bibinfo {author} {\bibfnamefont {A.}~\bibnamefont {Le~Tiec}}, \bibinfo {author} {\bibfnamefont {H.}~\bibnamefont {Nakano}}, \bibinfo {author} {\bibfnamefont {N.}~\bibnamefont {Sago}}, \ and\ \bibinfo {author} {\bibfnamefont {T.}~\bibnamefont {Tanaka}},\ }\href@noop {} {\bibfield  {journal} {\bibinfo  {journal} {Class. Quant. Grav.}\ }\textbf {\bibinfo {volume} {34}},\ \bibinfo {pages} {134001} (\bibinfo {year} {2017})}\BibitemShut {NoStop}%
\bibitem [{\citenamefont {Sch{\"a}fer}\ and\ \citenamefont {Jaranowski}(2018)}]{Schafer.al.18}%
  \BibitemOpen
  \bibfield  {author} {\bibinfo {author} {\bibfnamefont {G.}~\bibnamefont {Sch{\"a}fer}}\ and\ \bibinfo {author} {\bibfnamefont {P.}~\bibnamefont {Jaranowski}},\ }\href@noop {} {\bibfield  {journal} {\bibinfo  {journal} {Living Rev. Relativ.}\ }\textbf {\bibinfo {volume} {21}},\ \bibinfo {pages} {1} (\bibinfo {year} {2018})}\BibitemShut {NoStop}%
\bibitem [{\citenamefont {Poincar\'{e}}(2 99)}]{Poin}%
  \BibitemOpen
  \bibfield  {author} {\bibinfo {author} {\bibfnamefont {H.}~\bibnamefont {Poincar\'{e}}},\ }\href@noop {} {\emph {\bibinfo {title} {Les m\'{e}thodes nouvelles de la m\'{e}canique c\'{e}leste, Tome 1-3}}}\ (\bibinfo  {publisher} {Gauthiers-Vilard},\ \bibinfo {address} {Paris},\ \bibinfo {year} {1892-99})\BibitemShut {NoStop}%
\bibitem [{\citenamefont {Carter}(1968)}]{Carter.68}%
  \BibitemOpen
  \bibfield  {author} {\bibinfo {author} {\bibfnamefont {B.}~\bibnamefont {Carter}},\ }\href@noop {} {\bibfield  {journal} {\bibinfo  {journal} {Physical Review}\ }\textbf {\bibinfo {volume} {174}},\ \bibinfo {pages} {1559} (\bibinfo {year} {1968})}\BibitemShut {NoStop}%
\bibitem [{\citenamefont {Walker}\ and\ \citenamefont {Penrose}(1970)}]{WaPe.70}%
  \BibitemOpen
  \bibfield  {author} {\bibinfo {author} {\bibfnamefont {M.}~\bibnamefont {Walker}}\ and\ \bibinfo {author} {\bibfnamefont {R.}~\bibnamefont {Penrose}},\ }\href@noop {} {\bibfield  {journal} {\bibinfo  {journal} {Commun. Math. Phys.}\ }\textbf {\bibinfo {volume} {18}},\ \bibinfo {pages} {265} (\bibinfo {year} {1970})}\BibitemShut {NoStop}%
\bibitem [{\citenamefont {Floyd}(1973)}]{Floyd.73}%
  \BibitemOpen
  \bibfield  {author} {\bibinfo {author} {\bibfnamefont {R.}~\bibnamefont {Floyd}},\ }\href@noop {} {\bibfield  {journal} {\bibinfo  {journal} {PhD Thesis, University of London}\ } (\bibinfo {year} {1973})}\BibitemShut {NoStop}%
\bibitem [{\citenamefont {{Hughston}}\ and\ \citenamefont {{Sommers}}(1973)}]{HuSo.73}%
  \BibitemOpen
  \bibfield  {author} {\bibinfo {author} {\bibfnamefont {L.~P.}\ \bibnamefont {{Hughston}}}\ and\ \bibinfo {author} {\bibfnamefont {P.}~\bibnamefont {{Sommers}}},\ }\href@noop {} {\bibfield  {journal} {\bibinfo  {journal} {Commun. Math. Phys.}\ }\textbf {\bibinfo {volume} {33}},\ \bibinfo {pages} {129} (\bibinfo {year} {1973})}\BibitemShut {NoStop}%
\bibitem [{\citenamefont {Barausse}\ \emph {et~al.}(2009)\citenamefont {Barausse}, \citenamefont {Racine},\ and\ \citenamefont {Buonanno}}]{Ba.al.09}%
  \BibitemOpen
  \bibfield  {author} {\bibinfo {author} {\bibfnamefont {E.}~\bibnamefont {Barausse}}, \bibinfo {author} {\bibfnamefont {E.}~\bibnamefont {Racine}}, \ and\ \bibinfo {author} {\bibfnamefont {A.}~\bibnamefont {Buonanno}},\ }\href@noop {} {\bibfield  {journal} {\bibinfo  {journal} {Phys. Rev. D}\ }\textbf {\bibinfo {volume} {80}},\ \bibinfo {pages} {104025} (\bibinfo {year} {2009})},\ \bibinfo {note} {\textit{{E}rratum:} Phys. Rev. D \textbf{85}, 069904(E) (2012)}\BibitemShut {NoStop}%
\bibitem [{\citenamefont {Steinhoff}(2011)}]{Stei.11}%
  \BibitemOpen
  \bibfield  {author} {\bibinfo {author} {\bibfnamefont {J.}~\bibnamefont {Steinhoff}},\ }\href@noop {} {\bibfield  {journal} {\bibinfo  {journal} {Annalen der Physik}\ }\textbf {\bibinfo {volume} {523}},\ \bibinfo {pages} {296} (\bibinfo {year} {2011})}\BibitemShut {NoStop}%
\bibitem [{\citenamefont {d'Ambrosi}\ \emph {et~al.}(2015)\citenamefont {d'Ambrosi}, \citenamefont {Kumar},\ and\ \citenamefont {van Holten}}]{dAKuvHo.15}%
  \BibitemOpen
  \bibfield  {author} {\bibinfo {author} {\bibfnamefont {G.}~\bibnamefont {d'Ambrosi}}, \bibinfo {author} {\bibfnamefont {S.~S.}\ \bibnamefont {Kumar}}, \ and\ \bibinfo {author} {\bibfnamefont {J.}~\bibnamefont {van Holten}},\ }\href@noop {} {\bibfield  {journal} {\bibinfo  {journal} {Physics Letters B}\ }\textbf {\bibinfo {volume} {743}},\ \bibinfo {pages} {478} (\bibinfo {year} {2015})}\BibitemShut {NoStop}%
\bibitem [{\citenamefont {Vines}\ \emph {et~al.}(2016)\citenamefont {Vines}, \citenamefont {Kunst}, \citenamefont {Steinhoff},\ and\ \citenamefont {Hinderer}}]{ViKuStHi.16}%
  \BibitemOpen
  \bibfield  {author} {\bibinfo {author} {\bibfnamefont {J.}~\bibnamefont {Vines}}, \bibinfo {author} {\bibfnamefont {D.}~\bibnamefont {Kunst}}, \bibinfo {author} {\bibfnamefont {J.}~\bibnamefont {Steinhoff}}, \ and\ \bibinfo {author} {\bibfnamefont {T.}~\bibnamefont {Hinderer}},\ }\href@noop {} {\bibfield  {journal} {\bibinfo  {journal} {Phys. Rev. D}\ }\textbf {\bibinfo {volume} {93}},\ \bibinfo {pages} {103008} (\bibinfo {year} {2016})}\BibitemShut {NoStop}%
\bibitem [{\citenamefont {Witzany}\ \emph {et~al.}(2019)\citenamefont {Witzany}, \citenamefont {Steinhoff},\ and\ \citenamefont {Lukes-Gerakopoulos}}]{WiStLu.19}%
  \BibitemOpen
  \bibfield  {author} {\bibinfo {author} {\bibfnamefont {V.}~\bibnamefont {Witzany}}, \bibinfo {author} {\bibfnamefont {J.}~\bibnamefont {Steinhoff}}, \ and\ \bibinfo {author} {\bibfnamefont {G.}~\bibnamefont {Lukes-Gerakopoulos}},\ }\href@noop {} {\bibfield  {journal} {\bibinfo  {journal} {Class. Quant. Grav.}\ }\textbf {\bibinfo {volume} {36}},\ \bibinfo {pages} {075003} (\bibinfo {year} {2019})}\BibitemShut {NoStop}%
\bibitem [{\citenamefont {Kunst}\ \emph {et~al.}(2016)\citenamefont {Kunst}, \citenamefont {Ledvinka}, \citenamefont {Lukes-Gerakopoulos},\ and\ \citenamefont {Seyrich}}]{KuLeLuSe.16}%
  \BibitemOpen
  \bibfield  {author} {\bibinfo {author} {\bibfnamefont {D.}~\bibnamefont {Kunst}}, \bibinfo {author} {\bibfnamefont {T.}~\bibnamefont {Ledvinka}}, \bibinfo {author} {\bibfnamefont {G.}~\bibnamefont {Lukes-Gerakopoulos}}, \ and\ \bibinfo {author} {\bibfnamefont {J.}~\bibnamefont {Seyrich}},\ }\href@noop {} {\bibfield  {journal} {\bibinfo  {journal} {Phys. Rev. D}\ }\textbf {\bibinfo {volume} {93}},\ \bibinfo {pages} {044004} (\bibinfo {year} {2016})}\BibitemShut {NoStop}%
\bibitem [{\citenamefont {Witzany}(2019)}]{WitzHJ.19}%
  \BibitemOpen
  \bibfield  {author} {\bibinfo {author} {\bibfnamefont {V.}~\bibnamefont {Witzany}},\ }\href@noop {} {\bibfield  {journal} {\bibinfo  {journal} {Phys. Rev. D}\ }\textbf {\bibinfo {volume} {100}},\ \bibinfo {pages} {104030} (\bibinfo {year} {2019})}\BibitemShut {NoStop}%
\bibitem [{\citenamefont {Comp{\`e}re}\ and\ \citenamefont {Druart}(2022)}]{ComDru.22}%
  \BibitemOpen
  \bibfield  {author} {\bibinfo {author} {\bibfnamefont {G.}~\bibnamefont {Comp{\`e}re}}\ and\ \bibinfo {author} {\bibfnamefont {A.}~\bibnamefont {Druart}},\ }\href@noop {} {\bibfield  {journal} {\bibinfo  {journal} {SciPost Physics}\ }\textbf {\bibinfo {volume} {12}},\ \bibinfo {pages} {012} (\bibinfo {year} {2022})}\BibitemShut {NoStop}%
\bibitem [{\citenamefont {Grant}\ and\ \citenamefont {Moxon}(2023)}]{GrMo.23}%
  \BibitemOpen
  \bibfield  {author} {\bibinfo {author} {\bibfnamefont {A.~M.}\ \bibnamefont {Grant}}\ and\ \bibinfo {author} {\bibfnamefont {J.}~\bibnamefont {Moxon}},\ }\href@noop {} {\bibfield  {journal} {\bibinfo  {journal} {Phys. Rev. D}\ }\textbf {\bibinfo {volume} {108}},\ \bibinfo {pages} {104029} (\bibinfo {year} {2023})}\BibitemShut {NoStop}%
\bibitem [{\citenamefont {Lukes-Gerakopoulos}(2018)}]{Lukes.18}%
  \BibitemOpen
  \bibfield  {author} {\bibinfo {author} {\bibfnamefont {G.}~\bibnamefont {Lukes-Gerakopoulos}},\ }in\ \href@noop {} {\emph {\bibinfo {booktitle} {Proceedings of the MG14 Meeting on General Relativity, University of Rome “La Sapienza”, Italy, 12--18 July 2015}}}\ (\bibinfo {organization} {World Scientific},\ \bibinfo {year} {2018})\ pp.\ \bibinfo {pages} {1960--1965}\BibitemShut {NoStop}%
\bibitem [{\citenamefont {Marck}(1983)}]{Marck.83}%
  \BibitemOpen
  \bibfield  {author} {\bibinfo {author} {\bibfnamefont {J.-A.}\ \bibnamefont {Marck}},\ }\href@noop {} {\bibfield  {journal} {\bibinfo  {journal} {Proceedings of the Royal Society of London. A. Mathematical and Physical Sciences}\ }\textbf {\bibinfo {volume} {385}},\ \bibinfo {pages} {431} (\bibinfo {year} {1983})}\BibitemShut {NoStop}%
\bibitem [{\citenamefont {Van~de Meent}(2020)}]{VdM.20}%
  \BibitemOpen
  \bibfield  {author} {\bibinfo {author} {\bibfnamefont {M.}~\bibnamefont {Van~de Meent}},\ }\href@noop {} {\bibfield  {journal} {\bibinfo  {journal} {Class. Quant. Grav.}\ }\textbf {\bibinfo {volume} {37}},\ \bibinfo {pages} {145007} (\bibinfo {year} {2020})}\BibitemShut {NoStop}%
\bibitem [{\citenamefont {Wald}(1984)}]{Wald}%
  \BibitemOpen
  \bibfield  {author} {\bibinfo {author} {\bibfnamefont {R.~M.}\ \bibnamefont {Wald}},\ }\href@noop {} {\emph {\bibinfo {title} {General relativity}}}\ (\bibinfo  {publisher} {University of Chicago Press},\ \bibinfo {address} {Chicago},\ \bibinfo {year} {1984})\BibitemShut {NoStop}%
\bibitem [{\citenamefont {Arnold}\ \emph {et~al.}(2006)\citenamefont {Arnold}, \citenamefont {Kozlov}, \citenamefont {Neishtadt},\ and\ \citenamefont {Iacob}}]{Arn}%
  \BibitemOpen
  \bibfield  {author} {\bibinfo {author} {\bibfnamefont {V.~I.}\ \bibnamefont {Arnold}}, \bibinfo {author} {\bibfnamefont {V.~V.}\ \bibnamefont {Kozlov}}, \bibinfo {author} {\bibfnamefont {A.~I.}\ \bibnamefont {Neishtadt}}, \ and\ \bibinfo {author} {\bibfnamefont {I.}~\bibnamefont {Iacob}},\ }\href@noop {} {\emph {\bibinfo {title} {Mathematical aspects of classical and celestial mechanics}}},\ Vol.~\bibinfo {volume} {3}\ (\bibinfo  {publisher} {Springer},\ \bibinfo {year} {2006})\BibitemShut {NoStop}%
\bibitem [{\citenamefont {Ramond}(2024)}]{Ra.PapI.24}%
  \BibitemOpen
  \bibfield  {author} {\bibinfo {author} {\bibfnamefont {P.}~\bibnamefont {Ramond}},\ }\href@noop {} {\bibfield  {journal} {\bibinfo  {journal} {preprint}\ } (\bibinfo {year} {2024})},\ \Eprint {http://arxiv.org/abs/2210.03866} {arXiv:2210.03866} \BibitemShut {NoStop}%
\bibitem [{\citenamefont {Ramond}\ and\ \citenamefont {Druart}(2024)}]{Ra.al.quad.24}%
  \BibitemOpen
  \bibfield  {author} {\bibinfo {author} {\bibfnamefont {P.}~\bibnamefont {Ramond}}\ and\ \bibinfo {author} {\bibfnamefont {A.}~\bibnamefont {Druart}},\ }\href@noop {} {\bibfield  {journal} {\bibinfo  {journal} {unpublished}\ } (\bibinfo {year} {2024})}\BibitemShut {NoStop}%
\bibitem [{\citenamefont {Dixon}(1974)}]{Di.74}%
  \BibitemOpen
  \bibfield  {author} {\bibinfo {author} {\bibfnamefont {W.~G.}\ \bibnamefont {Dixon}},\ }\href@noop {} {\bibfield  {journal} {\bibinfo  {journal} {Phil. Trans. R. Soc. Lond. A}\ }\textbf {\bibinfo {volume} {277}},\ \bibinfo {pages} {59} (\bibinfo {year} {1974})}\BibitemShut {NoStop}%
\bibitem [{\citenamefont {Thorne}(1980)}]{Thorne.80}%
  \BibitemOpen
  \bibfield  {author} {\bibinfo {author} {\bibfnamefont {K.~S.}\ \bibnamefont {Thorne}},\ }\href@noop {} {\bibfield  {journal} {\bibinfo  {journal} {Reviews of Modern Physics}\ }\textbf {\bibinfo {volume} {52}},\ \bibinfo {pages} {299} (\bibinfo {year} {1980})}\BibitemShut {NoStop}%
\bibitem [{\citenamefont {Papapetrou}(1951)}]{Pa.51}%
  \BibitemOpen
  \bibfield  {author} {\bibinfo {author} {\bibfnamefont {A.}~\bibnamefont {Papapetrou}},\ }\href@noop {} {\bibfield  {journal} {\bibinfo  {journal} {Proc. R. Soc. Lond. A}\ }\textbf {\bibinfo {volume} {209}},\ \bibinfo {pages} {248} (\bibinfo {year} {1951})}\BibitemShut {NoStop}%
\bibitem [{\citenamefont {Tulczyjew}(1959)}]{Tu.59}%
  \BibitemOpen
  \bibfield  {author} {\bibinfo {author} {\bibfnamefont {W.}~\bibnamefont {Tulczyjew}},\ }\href@noop {} {\bibfield  {journal} {\bibinfo  {journal} {Acta Phys. Polon.}\ }\textbf {\bibinfo {volume} {18}},\ \bibinfo {pages} {393} (\bibinfo {year} {1959})}\BibitemShut {NoStop}%
\bibitem [{\citenamefont {Harte}(2012)}]{Ha.12}%
  \BibitemOpen
  \bibfield  {author} {\bibinfo {author} {\bibfnamefont {A.~I.}\ \bibnamefont {Harte}},\ }\href@noop {} {\bibfield  {journal} {\bibinfo  {journal} {Class. Quant. Grav.}\ }\textbf {\bibinfo {volume} {29}},\ \bibinfo {pages} {055012} (\bibinfo {year} {2012})}\BibitemShut {NoStop}%
\bibitem [{\citenamefont {{Harte}}(2020)}]{Ha.20}%
  \BibitemOpen
  \bibfield  {author} {\bibinfo {author} {\bibfnamefont {A.~I.}\ \bibnamefont {{Harte}}},\ }\href@noop {} {\bibfield  {journal} {\bibinfo  {journal} {Phys. Rev. D}\ }\textbf {\bibinfo {volume} {102}},\ \bibinfo {eid} {124075} (\bibinfo {year} {2020})}\BibitemShut {NoStop}%
\bibitem [{\citenamefont {Steinhoff}\ and\ \citenamefont {Puetzfeld}(2012)}]{StPu.12}%
  \BibitemOpen
  \bibfield  {author} {\bibinfo {author} {\bibfnamefont {J.}~\bibnamefont {Steinhoff}}\ and\ \bibinfo {author} {\bibfnamefont {D.}~\bibnamefont {Puetzfeld}},\ }\href@noop {} {\bibfield  {journal} {\bibinfo  {journal} {Phys. Rev. D}\ }\textbf {\bibinfo {volume} {86}},\ \bibinfo {pages} {044033} (\bibinfo {year} {2012})}\BibitemShut {NoStop}%
\bibitem [{\citenamefont {Marsat}(2015)}]{Ma.15}%
  \BibitemOpen
  \bibfield  {author} {\bibinfo {author} {\bibfnamefont {S.}~\bibnamefont {Marsat}},\ }\href@noop {} {\bibfield  {journal} {\bibinfo  {journal} {Class. Quant. Grav.}\ }\textbf {\bibinfo {volume} {32}},\ \bibinfo {pages} {085008} (\bibinfo {year} {2015})}\BibitemShut {NoStop}%
\bibitem [{\citenamefont {Marsat}(2015)}]{Ma.15}%
  \BibitemOpen
  \bibfield  {author} {\bibinfo {author} {\bibfnamefont {S.}~\bibnamefont {Marsat}},\ }\href@noop {} {\bibfield  {journal} {\bibinfo  {journal} {Class. Quant. Grav.}\ }\textbf {\bibinfo {volume} {32}},\ \bibinfo {pages} {085008} (\bibinfo {year} {2015})}\BibitemShut {NoStop}%
\bibitem [{\citenamefont {Binnington}\ and\ \citenamefont {Poisson}(2009)}]{BiPo.09}%
  \BibitemOpen
  \bibfield  {author} {\bibinfo {author} {\bibfnamefont {T.}~\bibnamefont {Binnington}}\ and\ \bibinfo {author} {\bibfnamefont {E.}~\bibnamefont {Poisson}},\ }\href@noop {} {\bibfield  {journal} {\bibinfo  {journal} {Phys. Rev. D}\ }\textbf {\bibinfo {volume} {80}},\ \bibinfo {pages} {084018} (\bibinfo {year} {2009})}\BibitemShut {NoStop}%
\bibitem [{\citenamefont {Laarakkers}\ and\ \citenamefont {Poisson}(1999)}]{LaPo.99}%
  \BibitemOpen
  \bibfield  {author} {\bibinfo {author} {\bibfnamefont {W.~G.}\ \bibnamefont {Laarakkers}}\ and\ \bibinfo {author} {\bibfnamefont {E.}~\bibnamefont {Poisson}},\ }\href@noop {} {\bibfield  {journal} {\bibinfo  {journal} {Astrophys. J.}\ }\textbf {\bibinfo {volume} {512}},\ \bibinfo {pages} {282} (\bibinfo {year} {1999})}\BibitemShut {NoStop}%
\bibitem [{\citenamefont {Boshkayev}\ \emph {et~al.}(2012)\citenamefont {Boshkayev}, \citenamefont {Rueda}, \citenamefont {Ruffini},\ and\ \citenamefont {Siutsou}}]{Bosh.al.12}%
  \BibitemOpen
  \bibfield  {author} {\bibinfo {author} {\bibfnamefont {K.}~\bibnamefont {Boshkayev}}, \bibinfo {author} {\bibfnamefont {J.~A.}\ \bibnamefont {Rueda}}, \bibinfo {author} {\bibfnamefont {R.}~\bibnamefont {Ruffini}}, \ and\ \bibinfo {author} {\bibfnamefont {I.}~\bibnamefont {Siutsou}},\ }\href@noop {} {\bibfield  {journal} {\bibinfo  {journal} {The Astrophysical Journal}\ }\textbf {\bibinfo {volume} {762}},\ \bibinfo {pages} {117} (\bibinfo {year} {2012})}\BibitemShut {NoStop}%
\bibitem [{\citenamefont {Uchikata}\ \emph {et~al.}(2016)\citenamefont {Uchikata}, \citenamefont {Yoshida},\ and\ \citenamefont {Pani}}]{Uchi.al.16}%
  \BibitemOpen
  \bibfield  {author} {\bibinfo {author} {\bibfnamefont {N.}~\bibnamefont {Uchikata}}, \bibinfo {author} {\bibfnamefont {S.}~\bibnamefont {Yoshida}}, \ and\ \bibinfo {author} {\bibfnamefont {P.}~\bibnamefont {Pani}},\ }\href@noop {} {\bibfield  {journal} {\bibinfo  {journal} {Physical Review D}\ }\textbf {\bibinfo {volume} {94}},\ \bibinfo {pages} {064015} (\bibinfo {year} {2016})}\BibitemShut {NoStop}%
\bibitem [{\citenamefont {Costa}\ and\ \citenamefont {Nat{\'a}rio}(2015)}]{CoNa.15}%
  \BibitemOpen
  \bibfield  {author} {\bibinfo {author} {\bibfnamefont {L.~F.~O.}\ \bibnamefont {Costa}}\ and\ \bibinfo {author} {\bibfnamefont {J.}~\bibnamefont {Nat{\'a}rio}},\ }\href@noop {} {\bibfield  {journal} {\bibinfo  {journal} {Fund. Theor. Phys.}\ }\textbf {\bibinfo {volume} {179}},\ \bibinfo {pages} {215} (\bibinfo {year} {2015})}\BibitemShut {NoStop}%
\bibitem [{\citenamefont {Gralla}\ \emph {et~al.}(2010)\citenamefont {Gralla}, \citenamefont {Harte},\ and\ \citenamefont {Wald}}]{GraHarWal.10}%
  \BibitemOpen
  \bibfield  {author} {\bibinfo {author} {\bibfnamefont {S.~E.}\ \bibnamefont {Gralla}}, \bibinfo {author} {\bibfnamefont {A.~I.}\ \bibnamefont {Harte}}, \ and\ \bibinfo {author} {\bibfnamefont {R.~M.}\ \bibnamefont {Wald}},\ }\href@noop {} {\bibfield  {journal} {\bibinfo  {journal} {Phys. Rev. D}\ }\textbf {\bibinfo {volume} {81}},\ \bibinfo {pages} {104012} (\bibinfo {year} {2010})}\BibitemShut {NoStop}%
\bibitem [{\citenamefont {Dixon}(1970)}]{DiI.70}%
  \BibitemOpen
  \bibfield  {author} {\bibinfo {author} {\bibfnamefont {W.~G.}\ \bibnamefont {Dixon}},\ }\href@noop {} {\bibfield  {journal} {\bibinfo  {journal} {Proc. R. Soc. Lond. A}\ }\textbf {\bibinfo {volume} {314}},\ \bibinfo {pages} {499–527} (\bibinfo {year} {1970})}\BibitemShut {NoStop}%
\bibitem [{\citenamefont {Compère}\ \emph {et~al.}(2023)\citenamefont {Compère}, \citenamefont {Druart},\ and\ \citenamefont {Vines}}]{ComDruVin.23}%
  \BibitemOpen
  \bibfield  {author} {\bibinfo {author} {\bibfnamefont {G.}~\bibnamefont {Compère}}, \bibinfo {author} {\bibfnamefont {A.}~\bibnamefont {Druart}}, \ and\ \bibinfo {author} {\bibfnamefont {J.}~\bibnamefont {Vines}},\ }\href@noop {} {\bibfield  {journal} {\bibinfo  {journal} {SciPost Phys.}\ }\textbf {\bibinfo {volume} {15}},\ \bibinfo {pages} {226} (\bibinfo {year} {2023})}\BibitemShut {NoStop}%
\bibitem [{\citenamefont {Souriau}(1970)}]{Souri.70}%
  \BibitemOpen
  \bibfield  {author} {\bibinfo {author} {\bibfnamefont {J.-M.}\ \bibnamefont {Souriau}},\ }\href@noop {} {\bibfield  {journal} {\bibinfo  {journal} {CR Acad. Sci. Paris}\ }\textbf {\bibinfo {volume} {271}},\ \bibinfo {pages} {751} (\bibinfo {year} {1970})}\BibitemShut {NoStop}%
\bibitem [{\citenamefont {K{\"u}nzle}(1972)}]{Kun.72}%
  \BibitemOpen
  \bibfield  {author} {\bibinfo {author} {\bibfnamefont {H.}~\bibnamefont {K{\"u}nzle}},\ }\href@noop {} {\bibfield  {journal} {\bibinfo  {journal} {Journal of Mathematical Physics}\ }\textbf {\bibinfo {volume} {13}},\ \bibinfo {pages} {739} (\bibinfo {year} {1972})}\BibitemShut {NoStop}%
\bibitem [{\citenamefont {Vaisman}(2012)}]{Vaisman.12}%
  \BibitemOpen
  \bibfield  {author} {\bibinfo {author} {\bibfnamefont {I.}~\bibnamefont {Vaisman}},\ }\href@noop {} {\emph {\bibinfo {title} {Lectures on the geometry of Poisson manifolds}}},\ Vol.\ \bibinfo {volume} {118}\ (\bibinfo  {publisher} {Birkh{\"a}user},\ \bibinfo {year} {2012})\BibitemShut {NoStop}%
\bibitem [{\citenamefont {Deriglazov}(2022)}]{Derigl.22}%
  \BibitemOpen
  \bibfield  {author} {\bibinfo {author} {\bibfnamefont {A.~A.}\ \bibnamefont {Deriglazov}},\ }\href@noop {} {\bibfield  {journal} {\bibinfo  {journal} {Universe}\ }\textbf {\bibinfo {volume} {8}},\ \bibinfo {pages} {536} (\bibinfo {year} {2022})}\BibitemShut {NoStop}%
\bibitem [{\citenamefont {Ramond}\ and\ \citenamefont {{Isoyama}}(2024)}]{Ra.Iso.PapII.24}%
  \BibitemOpen
  \bibfield  {author} {\bibinfo {author} {\bibfnamefont {P.}~\bibnamefont {Ramond}}\ and\ \bibinfo {author} {\bibfnamefont {S.}~\bibnamefont {{Isoyama}}},\ }\href@noop {} {\  (\bibinfo {year} {2024})}\BibitemShut {NoStop}%
\bibitem [{\citenamefont {Zambon}(2011)}]{Zambon.11}%
  \BibitemOpen
  \bibfield  {author} {\bibinfo {author} {\bibfnamefont {M.}~\bibnamefont {Zambon}},\ }in\ \href@noop {} {\emph {\bibinfo {booktitle} {Complex and Differential Geometry: Conference held at Leibniz Universit{\"a}t Hannover, September 14--18, 2009}}}\ (\bibinfo {organization} {Springer},\ \bibinfo {year} {2011})\ pp.\ \bibinfo {pages} {403--420}\BibitemShut {NoStop}%
\bibitem [{\citenamefont {Bursztyn}(2013)}]{Burs.13}%
  \BibitemOpen
  \bibfield  {author} {\bibinfo {author} {\bibfnamefont {H.}~\bibnamefont {Bursztyn}},\ }\href@noop {} {\bibfield  {journal} {\bibinfo  {journal} {Geometric and topological methods for quantum field theory}\ ,\ \bibinfo {pages} {4}} (\bibinfo {year} {2013})}\BibitemShut {NoStop}%
\bibitem [{\citenamefont {Liouville}(1855)}]{Liou1855}%
  \BibitemOpen
  \bibfield  {author} {\bibinfo {author} {\bibfnamefont {J.}~\bibnamefont {Liouville}},\ }\href@noop {} {\bibfield  {journal} {\bibinfo  {journal} {Journal de Math{\'e}matiques pures et appliqu{\'e}es}\ }\textbf {\bibinfo {volume} {20}},\ \bibinfo {pages} {137} (\bibinfo {year} {1855})}\BibitemShut {NoStop}%
\bibitem [{\citenamefont {Hinderer}\ and\ \citenamefont {Flanagan}(2008)}]{HiFl.08}%
  \BibitemOpen
  \bibfield  {author} {\bibinfo {author} {\bibfnamefont {T.}~\bibnamefont {Hinderer}}\ and\ \bibinfo {author} {\bibfnamefont {{\'E}.}~\bibnamefont {Flanagan}},\ }\href@noop {} {\bibfield  {journal} {\bibinfo  {journal} {Phys. Rev. D}\ }\textbf {\bibinfo {volume} {78}},\ \bibinfo {pages} {064028} (\bibinfo {year} {2008})}\BibitemShut {NoStop}%
\bibitem [{\citenamefont {C{\'a}rdenas-Avenda{\~n}o}\ \emph {et~al.}(2018)\citenamefont {C{\'a}rdenas-Avenda{\~n}o}, \citenamefont {Gutierrez}, \citenamefont {Pach{\'o}n},\ and\ \citenamefont {Yunes}}]{Card.al.18}%
  \BibitemOpen
  \bibfield  {author} {\bibinfo {author} {\bibfnamefont {A.}~\bibnamefont {C{\'a}rdenas-Avenda{\~n}o}}, \bibinfo {author} {\bibfnamefont {A.~F.}\ \bibnamefont {Gutierrez}}, \bibinfo {author} {\bibfnamefont {L.~A.}\ \bibnamefont {Pach{\'o}n}}, \ and\ \bibinfo {author} {\bibfnamefont {N.}~\bibnamefont {Yunes}},\ }\href@noop {} {\bibfield  {journal} {\bibinfo  {journal} {Classical and Quantum Gravity}\ }\textbf {\bibinfo {volume} {35}},\ \bibinfo {pages} {165010} (\bibinfo {year} {2018})}\BibitemShut {NoStop}%
\bibitem [{\citenamefont {Tanay}\ \emph {et~al.}(2021)\citenamefont {Tanay}, \citenamefont {Stein},\ and\ \citenamefont {Ghersi}}]{Tanay.al.21}%
  \BibitemOpen
  \bibfield  {author} {\bibinfo {author} {\bibfnamefont {S.}~\bibnamefont {Tanay}}, \bibinfo {author} {\bibfnamefont {L.~C.}\ \bibnamefont {Stein}}, \ and\ \bibinfo {author} {\bibfnamefont {J.~T.~G.}\ \bibnamefont {Ghersi}},\ }\href@noop {} {\bibfield  {journal} {\bibinfo  {journal} {Physical Review D}\ }\textbf {\bibinfo {volume} {103}},\ \bibinfo {pages} {064066} (\bibinfo {year} {2021})}\BibitemShut {NoStop}%
\bibitem [{\citenamefont {Fasano}\ and\ \citenamefont {Marmi}(2006)}]{FaMa.06}%
  \BibitemOpen
  \bibfield  {author} {\bibinfo {author} {\bibfnamefont {A.}~\bibnamefont {Fasano}}\ and\ \bibinfo {author} {\bibfnamefont {S.}~\bibnamefont {Marmi}},\ }\href@noop {} {\emph {\bibinfo {title} {Analytical mechanics: an introduction}}}\ (\bibinfo  {publisher} {Oxford University Press},\ \bibinfo {year} {2006})\BibitemShut {NoStop}%
\bibitem [{\citenamefont {R{\"u}diger}(1981)}]{Rudiger.I.81}%
  \BibitemOpen
  \bibfield  {author} {\bibinfo {author} {\bibfnamefont {R.}~\bibnamefont {R{\"u}diger}},\ }\href@noop {} {\bibfield  {journal} {\bibinfo  {journal} {Proceedings of the Royal Society of London. A. Mathematical and Physical Sciences}\ }\textbf {\bibinfo {volume} {375}},\ \bibinfo {pages} {185} (\bibinfo {year} {1981})}\BibitemShut {NoStop}%
\bibitem [{\citenamefont {R{\"u}diger}(1983)}]{Rudiger.II.83}%
  \BibitemOpen
  \bibfield  {author} {\bibinfo {author} {\bibfnamefont {R.}~\bibnamefont {R{\"u}diger}},\ }\href@noop {} {\bibfield  {journal} {\bibinfo  {journal} {Proceedings of the Royal Society of London. A. Mathematical and Physical Sciences}\ }\textbf {\bibinfo {volume} {385}},\ \bibinfo {pages} {229} (\bibinfo {year} {1983})}\BibitemShut {NoStop}%
\bibitem [{\citenamefont {Gourgoulhon}(2021)}]{GourgoulhonBH}%
  \BibitemOpen
  \bibfield  {author} {\bibinfo {author} {\bibfnamefont {{\'E}.}~\bibnamefont {Gourgoulhon}},\ }\href {https://relativite.obspm.fr/blackholes/} {\enquote {\bibinfo {title} {Geometry and physics of black holes},}\ } (\bibinfo {year} {2021})\BibitemShut {NoStop}%
\bibitem [{MMA(2024)}]{MMAPRL}%
  \BibitemOpen
  \href@noop {} {\enquote {\bibinfo {title} {Companion Mathematica Notebook},}\ } (\bibinfo {year} {2024})\BibitemShut {NoStop}%
\bibitem [{\citenamefont {Pound}\ and\ \citenamefont {Wardell}(2021)}]{PoWa.21}%
  \BibitemOpen
  \bibfield  {author} {\bibinfo {author} {\bibfnamefont {A.}~\bibnamefont {Pound}}\ and\ \bibinfo {author} {\bibfnamefont {B.}~\bibnamefont {Wardell}},\ }\href@noop {} {\enquote {\bibinfo {title} {Black hole perturbation theory and gravitational self-force},}\ } (\bibinfo {year} {2021}),\ \bibinfo {note} {invited chapter for "Handbook of Gravitational Wave Astronomy" (Eds. C. Bambi, S. Katsanevas, and K. Kokkotas; Springer, Singapore, 2021)}\BibitemShut {NoStop}%
\bibitem [{\citenamefont {Afshordi}\ \emph {et~al.}(2023)\citenamefont {Afshordi}, \citenamefont {Ak{\c{c}}ay}, \citenamefont {Seoane}, \citenamefont {Antonelli}, \citenamefont {Aurrekoetxea}, \citenamefont {Barack}, \citenamefont {Barausse}, \citenamefont {Benkel}, \citenamefont {Bernard}, \citenamefont {Bernuzzi} \emph {et~al.}}]{LISAWHITE.23}%
  \BibitemOpen
  \bibfield  {author} {\bibinfo {author} {\bibfnamefont {N.}~\bibnamefont {Afshordi}}, \bibinfo {author} {\bibfnamefont {S.}~\bibnamefont {Ak{\c{c}}ay}}, \bibinfo {author} {\bibfnamefont {P.~A.}\ \bibnamefont {Seoane}}, \bibinfo {author} {\bibfnamefont {A.}~\bibnamefont {Antonelli}}, \bibinfo {author} {\bibfnamefont {J.~C.}\ \bibnamefont {Aurrekoetxea}}, \bibinfo {author} {\bibfnamefont {L.}~\bibnamefont {Barack}}, \bibinfo {author} {\bibfnamefont {E.}~\bibnamefont {Barausse}}, \bibinfo {author} {\bibfnamefont {R.}~\bibnamefont {Benkel}}, \bibinfo {author} {\bibfnamefont {L.}~\bibnamefont {Bernard}}, \bibinfo {author} {\bibfnamefont {S.}~\bibnamefont {Bernuzzi}},  \emph {et~al.},\ }\href@noop {} {\bibfield  {journal} {\bibinfo  {journal} {arXiv preprint arXiv:2311.01300}\ } (\bibinfo {year} {2023})}\BibitemShut {NoStop}%
\bibitem [{\citenamefont {Schmidt}(2002)}]{Schm.02}%
  \BibitemOpen
  \bibfield  {author} {\bibinfo {author} {\bibfnamefont {W.}~\bibnamefont {Schmidt}},\ }\href@noop {} {\bibfield  {journal} {\bibinfo  {journal} {Class. Quant. Grav.}\ }\textbf {\bibinfo {volume} {19}},\ \bibinfo {pages} {2743} (\bibinfo {year} {2002})}\BibitemShut {NoStop}%
\bibitem [{\citenamefont {Drummond}\ and\ \citenamefont {Hughes}(2022{\natexlab{a}})}]{DruHug.I.22}%
  \BibitemOpen
  \bibfield  {author} {\bibinfo {author} {\bibfnamefont {L.~V.}\ \bibnamefont {Drummond}}\ and\ \bibinfo {author} {\bibfnamefont {S.~A.}\ \bibnamefont {Hughes}},\ }\href@noop {} {\bibfield  {journal} {\bibinfo  {journal} {Phys. Rev. D}\ }\textbf {\bibinfo {volume} {105}},\ \bibinfo {pages} {124040} (\bibinfo {year} {2022}{\natexlab{a}})}\BibitemShut {NoStop}%
\bibitem [{\citenamefont {Drummond}\ and\ \citenamefont {Hughes}(2022{\natexlab{b}})}]{DruHug.II.22}%
  \BibitemOpen
  \bibfield  {author} {\bibinfo {author} {\bibfnamefont {L.~V.}\ \bibnamefont {Drummond}}\ and\ \bibinfo {author} {\bibfnamefont {S.~A.}\ \bibnamefont {Hughes}},\ }\href@noop {} {\bibfield  {journal} {\bibinfo  {journal} {Phys. Rev. D}\ }\textbf {\bibinfo {volume} {105}},\ \bibinfo {pages} {124041} (\bibinfo {year} {2022}{\natexlab{b}})}\BibitemShut {NoStop}%
\bibitem [{\citenamefont {Isoyama}\ \emph {et~al.}(2019)\citenamefont {Isoyama}, \citenamefont {Fujita}, \citenamefont {Nakano}, \citenamefont {Sago},\ and\ \citenamefont {Tanaka}}]{IsoAl.19}%
  \BibitemOpen
  \bibfield  {author} {\bibinfo {author} {\bibfnamefont {S.}~\bibnamefont {Isoyama}}, \bibinfo {author} {\bibfnamefont {R.}~\bibnamefont {Fujita}}, \bibinfo {author} {\bibfnamefont {H.}~\bibnamefont {Nakano}}, \bibinfo {author} {\bibfnamefont {N.}~\bibnamefont {Sago}}, \ and\ \bibinfo {author} {\bibfnamefont {T.}~\bibnamefont {Tanaka}},\ }\href@noop {} {\bibfield  {journal} {\bibinfo  {journal} {Prog. Theor. Exp. Phys.}\ }\textbf {\bibinfo {volume} {2019}},\ \bibinfo {pages} {013E01} (\bibinfo {year} {2019})}\BibitemShut {NoStop}%
\bibitem [{\citenamefont {Isoyama}\ \emph {et~al.}(2014)\citenamefont {Isoyama} \emph {et~al.}}]{Is.al.14}%
  \BibitemOpen
  \bibfield  {author} {\bibinfo {author} {\bibfnamefont {S.}~\bibnamefont {Isoyama}} \emph {et~al.},\ }\href@noop {} {\bibfield  {journal} {\bibinfo  {journal} {Phys. Rev. Lett.}\ }\textbf {\bibinfo {volume} {113}},\ \bibinfo {pages} {161101} (\bibinfo {year} {2014})}\BibitemShut {NoStop}%
\bibitem [{\citenamefont {{Le Tiec}}(2015)}]{Le.15}%
  \BibitemOpen
  \bibfield  {author} {\bibinfo {author} {\bibfnamefont {A.}~\bibnamefont {{Le Tiec}}},\ }\href@noop {} {\bibfield  {journal} {\bibinfo  {journal} {Phys. Rev. D}\ }\textbf {\bibinfo {volume} {92}},\ \bibinfo {pages} {084021} (\bibinfo {year} {2015})}\BibitemShut {NoStop}%
\bibitem [{\citenamefont {Pound}\ \emph {et~al.}(2020)\citenamefont {Pound}, \citenamefont {Wardell}, \citenamefont {Warburton},\ and\ \citenamefont {Miller}}]{Po.al.20}%
  \BibitemOpen
  \bibfield  {author} {\bibinfo {author} {\bibfnamefont {A.}~\bibnamefont {Pound}}, \bibinfo {author} {\bibfnamefont {B.}~\bibnamefont {Wardell}}, \bibinfo {author} {\bibfnamefont {N.}~\bibnamefont {Warburton}}, \ and\ \bibinfo {author} {\bibfnamefont {J.}~\bibnamefont {Miller}},\ }\href@noop {} {\bibfield  {journal} {\bibinfo  {journal} {Phys. Rev. Lett.}\ }\textbf {\bibinfo {volume} {124}},\ \bibinfo {pages} {021101} (\bibinfo {year} {2020})}\BibitemShut {NoStop}%
\bibitem [{\citenamefont {{Le Tiec}}(2014)}]{Le2.14}%
  \BibitemOpen
  \bibfield  {author} {\bibinfo {author} {\bibfnamefont {A.}~\bibnamefont {{Le Tiec}}},\ }\href@noop {} {\bibfield  {journal} {\bibinfo  {journal} {Int. J. Mod. Phys. D}\ }\textbf {\bibinfo {volume} {23}},\ \bibinfo {pages} {1430022} (\bibinfo {year} {2014})}\BibitemShut {NoStop}%
\bibitem [{\citenamefont {Blanco}\ and\ \citenamefont {Flanagan}(2023{\natexlab{a}})}]{BlFl2.23}%
  \BibitemOpen
  \bibfield  {author} {\bibinfo {author} {\bibfnamefont {F.}~\bibnamefont {Blanco}}\ and\ \bibinfo {author} {\bibfnamefont {E.~.}\ \bibnamefont {Flanagan}},\ }\href@noop {} {\bibfield  {journal} {\bibinfo  {journal} {Phys. Rev. D}\ }\textbf {\bibinfo {volume} {107}},\ \bibinfo {pages} {124017} (\bibinfo {year} {2023}{\natexlab{a}})}\BibitemShut {NoStop}%
\bibitem [{\citenamefont {Galley}(2013)}]{Galley.13}%
  \BibitemOpen
  \bibfield  {author} {\bibinfo {author} {\bibfnamefont {C.~R.}\ \bibnamefont {Galley}},\ }\href@noop {} {\bibfield  {journal} {\bibinfo  {journal} {Phys. Rev. Lett.}\ }\textbf {\bibinfo {volume} {110}},\ \bibinfo {pages} {174301} (\bibinfo {year} {2013})}\BibitemShut {NoStop}%
\bibitem [{\citenamefont {Nasipak}(2022)}]{Nasipak.22}%
  \BibitemOpen
  \bibfield  {author} {\bibinfo {author} {\bibfnamefont {Z.}~\bibnamefont {Nasipak}},\ }\href@noop {} {\bibfield  {journal} {\bibinfo  {journal} {Phys. Rev. D}\ }\textbf {\bibinfo {volume} {106}},\ \bibinfo {pages} {064042} (\bibinfo {year} {2022})}\BibitemShut {NoStop}%
\bibitem [{\citenamefont {Blanco}\ and\ \citenamefont {Flanagan}(2023{\natexlab{b}})}]{BlFl.23}%
  \BibitemOpen
  \bibfield  {author} {\bibinfo {author} {\bibfnamefont {F.}~\bibnamefont {Blanco}}\ and\ \bibinfo {author} {\bibfnamefont {E.}~\bibnamefont {Flanagan}},\ }\href@noop {} {\bibfield  {journal} {\bibinfo  {journal} {Phys. Rev. Lett.}\ }\textbf {\bibinfo {volume} {130}},\ \bibinfo {pages} {051201} (\bibinfo {year} {2023}{\natexlab{b}})}\BibitemShut {NoStop}%
\bibitem [{\citenamefont {Steinhoff}\ and\ \citenamefont {Puetzfeld}(2010)}]{StPu.10}%
  \BibitemOpen
  \bibfield  {author} {\bibinfo {author} {\bibfnamefont {J.}~\bibnamefont {Steinhoff}}\ and\ \bibinfo {author} {\bibfnamefont {D.}~\bibnamefont {Puetzfeld}},\ }\href@noop {} {\bibfield  {journal} {\bibinfo  {journal} {Phys. Rev. D}\ }\textbf {\bibinfo {volume} {81}},\ \bibinfo {pages} {044019} (\bibinfo {year} {2010})}\BibitemShut {NoStop}%
\bibitem [{\citenamefont {Ramond}\ and\ \citenamefont {{Le Tiec}}(2021)}]{RaLe.20}%
  \BibitemOpen
  \bibfield  {author} {\bibinfo {author} {\bibfnamefont {P.}~\bibnamefont {Ramond}}\ and\ \bibinfo {author} {\bibfnamefont {A.}~\bibnamefont {{Le Tiec}}},\ }\href@noop {} {\bibfield  {journal} {\bibinfo  {journal} {Class. Quant. Grav}\ }\textbf {\bibinfo {volume} {38}} (\bibinfo {year} {2021})}\BibitemShut {NoStop}%
\bibitem [{\citenamefont {Frolov}\ \emph {et~al.}(2017)\citenamefont {Frolov}, \citenamefont {Krtou{\v{s}}},\ and\ \citenamefont {Kubiz{\v{n}}{\'a}k}}]{Frolov.17}%
  \BibitemOpen
  \bibfield  {author} {\bibinfo {author} {\bibfnamefont {V.~P.}\ \bibnamefont {Frolov}}, \bibinfo {author} {\bibfnamefont {P.}~\bibnamefont {Krtou{\v{s}}}}, \ and\ \bibinfo {author} {\bibfnamefont {D.}~\bibnamefont {Kubiz{\v{n}}{\'a}k}},\ }\href@noop {} {\bibfield  {journal} {\bibinfo  {journal} {Living Rev. Relativ.}\ }\textbf {\bibinfo {volume} {20}},\ \bibinfo {pages} {1} (\bibinfo {year} {2017})}\BibitemShut {NoStop}%
\bibitem [{\citenamefont {Zhong}\ \emph {et~al.}(2010)\citenamefont {Zhong}, \citenamefont {Wu}, \citenamefont {Liu},\ and\ \citenamefont {Deng}}]{Zho.al.10}%
  \BibitemOpen
  \bibfield  {author} {\bibinfo {author} {\bibfnamefont {S.-Y.}\ \bibnamefont {Zhong}}, \bibinfo {author} {\bibfnamefont {X.}~\bibnamefont {Wu}}, \bibinfo {author} {\bibfnamefont {S.-Q.}\ \bibnamefont {Liu}}, \ and\ \bibinfo {author} {\bibfnamefont {X.-F.}\ \bibnamefont {Deng}},\ }\href@noop {} {\bibfield  {journal} {\bibinfo  {journal} {Physical Review D}\ }\textbf {\bibinfo {volume} {82}},\ \bibinfo {pages} {124040} (\bibinfo {year} {2010})}\BibitemShut {NoStop}%
\bibitem [{\citenamefont {Seyrich}\ and\ \citenamefont {Lukes-Gerakopoulos}(2012)}]{Sey.Lu.12}%
  \BibitemOpen
  \bibfield  {author} {\bibinfo {author} {\bibfnamefont {J.}~\bibnamefont {Seyrich}}\ and\ \bibinfo {author} {\bibfnamefont {G.}~\bibnamefont {Lukes-Gerakopoulos}},\ }\href@noop {} {\bibfield  {journal} {\bibinfo  {journal} {Physical Review D}\ }\textbf {\bibinfo {volume} {86}},\ \bibinfo {pages} {124013} (\bibinfo {year} {2012})}\BibitemShut {NoStop}%
\bibitem [{\citenamefont {Wu}\ \emph {et~al.}(2021)\citenamefont {Wu}, \citenamefont {Wang}, \citenamefont {Sun},\ and\ \citenamefont {Liu}}]{Wu.Kerr.21}%
  \BibitemOpen
  \bibfield  {author} {\bibinfo {author} {\bibfnamefont {X.}~\bibnamefont {Wu}}, \bibinfo {author} {\bibfnamefont {Y.}~\bibnamefont {Wang}}, \bibinfo {author} {\bibfnamefont {W.}~\bibnamefont {Sun}}, \ and\ \bibinfo {author} {\bibfnamefont {F.}~\bibnamefont {Liu}},\ }\href@noop {} {\bibfield  {journal} {\bibinfo  {journal} {The Astrophysical Journal}\ }\textbf {\bibinfo {volume} {914}},\ \bibinfo {pages} {63} (\bibinfo {year} {2021})}\BibitemShut {NoStop}%
\bibitem [{\citenamefont {Brink}\ \emph {et~al.}(2015)\citenamefont {Brink}, \citenamefont {Geyer},\ and\ \citenamefont {Hinderer}}]{BrGeHi.15}%
  \BibitemOpen
  \bibfield  {author} {\bibinfo {author} {\bibfnamefont {J.}~\bibnamefont {Brink}}, \bibinfo {author} {\bibfnamefont {M.}~\bibnamefont {Geyer}}, \ and\ \bibinfo {author} {\bibfnamefont {T.}~\bibnamefont {Hinderer}},\ }\href@noop {} {\bibfield  {journal} {\bibinfo  {journal} {Phys. Rev. D}\ }\textbf {\bibinfo {volume} {91}},\ \bibinfo {pages} {083001} (\bibinfo {year} {2015})}\BibitemShut {NoStop}%
\bibitem [{\citenamefont {Mukherjee}\ and\ \citenamefont {Tripathy}(2020)}]{MukTri.20}%
  \BibitemOpen
  \bibfield  {author} {\bibinfo {author} {\bibfnamefont {S.}~\bibnamefont {Mukherjee}}\ and\ \bibinfo {author} {\bibfnamefont {S.}~\bibnamefont {Tripathy}},\ }\href@noop {} {\bibfield  {journal} {\bibinfo  {journal} {Phys. Rev. D}\ }\textbf {\bibinfo {volume} {101}},\ \bibinfo {pages} {124047} (\bibinfo {year} {2020})}\BibitemShut {NoStop}%
\bibitem [{\citenamefont {Ruangsri}\ and\ \citenamefont {Hughes}(2014)}]{RuHu.14}%
  \BibitemOpen
  \bibfield  {author} {\bibinfo {author} {\bibfnamefont {U.}~\bibnamefont {Ruangsri}}\ and\ \bibinfo {author} {\bibfnamefont {S.~A.}\ \bibnamefont {Hughes}},\ }\href@noop {} {\bibfield  {journal} {\bibinfo  {journal} {Phys. Rev. D}\ }\textbf {\bibinfo {volume} {89}},\ \bibinfo {pages} {084036} (\bibinfo {year} {2014})}\BibitemShut {NoStop}%
\bibitem [{\citenamefont {Maggiore}\ \emph {et~al.}(2020)\citenamefont {Maggiore} \emph {et~al.}}]{ET.al.20}%
  \BibitemOpen
  \bibfield  {author} {\bibinfo {author} {\bibfnamefont {M.}~\bibnamefont {Maggiore}} \emph {et~al.},\ }\href@noop {} {\bibfield  {journal} {\bibinfo  {journal} {J. Cosmol. Astropart. P.}\ }\textbf {\bibinfo {volume} {2020}},\ \bibinfo {pages} {050} (\bibinfo {year} {2020})}\BibitemShut {NoStop}%
\bibitem [{\citenamefont {Hall}(2022)}]{CE.22}%
  \BibitemOpen
  \bibfield  {author} {\bibinfo {author} {\bibfnamefont {E.~D.}\ \bibnamefont {Hall}},\ }\href@noop {} {\bibfield  {journal} {\bibinfo  {journal} {Galaxies}\ }\textbf {\bibinfo {volume} {10}},\ \bibinfo {pages} {90} (\bibinfo {year} {2022})}\BibitemShut {NoStop}%
\end{thebibliography}


\end{document}